\documentclass[twocolumn,showpacs,showkeys]{revtex4}
\pdfoutput=1    
\usepackage{graphicx,amsmath}

\newcommand{\p}{\partial}
\newcommand{\rp}{\right)}
\newcommand{\lp}{\left(}
\newcommand{\rb}{\right]}
\newcommand{\lb}{\left[}

\newcommand{\al}{\alpha}

\newcommand{\ga}{\gamma}
\newcommand{\de}{\delta}
\newcommand{\ep}{\epsilon}
\newcommand{\ze}{\zeta}
\newcommand{\ta}{\theta}
\newcommand{\ka}{\kappa}
\newcommand{\la}{\lambda}
\newcommand{\om}{\omega}
\newcommand{\si}{\sigma}
\newcommand{\De}{\Delta}
\newcommand{\Ga}{\Gamma}
\newcommand{\La}{\Lambda}
\newcommand{\Om}{\Omega}
\newcommand{\Si}{\Sigma}

\newcommand{\atan}{\mathrm{atan}} 
\newcommand{\bmin}{b_\mathrm{min}} 
\newcommand{\dep}{\textsc{deplete}}
\newcommand{\colonl}{(Color online.)}
\newcommand{\erfc}{\mathrm{erfc}} \newcommand{\ftd}{p\textsc{f3d}}
\newcommand{\hydra}{\textsc{hydra}}
\newcommand{\im}{\mathrm{Im}} 
\newcommand{\ioot}{\ensuremath{i_1^\mathrm{OT}}}
\newcommand{\lip}{\textsc{newlip}}
\newcommand{\las}{\textsc{lasnex}} \newcommand{\slip}{\textsc{slip}}
\newcommand{\npnd}{n_\mathrm{pnd}} \newcommand{\re}{\mathrm{Re}}
\newcommand{\sumjion}{\sum_{j\in\mathrm{ions}}}

\begin{document}

\title{Ray-based calculations of backscatter in laser fusion targets} 
\author{D.\ J.\ Strozzi}
\email{strozzi2@llnl.gov}
\author{E.\ A.\ Williams}
\author{D.\ E.\ Hinkel} 
\author{D.\ H.\ Froula} 
\author{R.\ A.\ London} 
\author{D.\ A.\ Callahan} 

\affiliation{AX Division, Lawrence Livermore National Laboratory,
7000 East Avenue, Livermore, CA 94550}


\date{\today}

\begin{abstract}
  A 1D, steady-state model for Brillouin and Raman backscatter from an
  inhomogeneous plasma is presented. The daughter plasma waves are
  treated in the strong damping limit, and have amplitudes given by
  the (linear) kinetic response to the ponderomotive drive. Pump
  depletion, inverse-bremsstrahlung damping, bremsstrahlung emission,
  Thomson scattering off density fluctuations, and whole-beam focusing
  are included. The numerical code \dep{}, which implements this
  model, is described. The model is compared with traditional linear
  gain calculations, as well as ``plane-wave'' simulations with the
  paraxial propagation code \ftd{}. Comparisons with
  Brillouin-scattering experiments at the OMEGA Laser Facility [T.\
  R.\ Boehly et al., Opt.\ Commun. 133, p.\ 495 (1997)] show that
  laser speckles greatly enhance the reflectivity over the \dep{}
  results. An approximate upper bound on this enhancement, motivated
  by phase conjugation, is given by doubling the \dep{} coupling
  coefficient. Analysis with \dep{} of an ignition design for the
  National Ignition Facility (NIF) [J.\ A.\ Paisner, E.\ M.\ Campbell,
  and W.\ J.\ Hogan, Fusion Technol.\ 26, p.\ 755 (1994)], with a peak
  radiation temperature of 285 eV, shows encouragingly low
  reflectivity. Re-absorption of Raman light is seen to be significant
  in this design.
\end{abstract}
\maketitle

\section{Introduction}
Laser-plasma interaction (LPI) \cite{kruer-lpi-1988} is an important
plasma-physics problem which poses serious challenges to theoretical
modeling.  LPI is the basis of several applications, including
laser-based particle acceleration \cite{tajima-laseraccel-prl-1979}
and the backward Raman amplifier \cite{malkin-ramanamp-prl-1999}.
Moreover, for inertial confinement fusion (ICF)\cite{atzeni-icf-2004,lindl-nif-pop-2004} to succeed, LPI must
not be so active that it prevents the desired laser energy from being
delivered to the target, with the desired spatial and temporal
behavior. This paper focuses on modeling the backscatter
instabilities, where a laser light wave (mode 0) decays into a
backscattered light wave (mode 1) and a plasma wave (mode 2). In
stimulated Raman scattering (SRS) and stimulated Brillouin scattering
(SBS), the plasma wave is, respectively, an electron plasma wave and
an ion acoustic wave. These LPI processes pose a serious risk to indirect-drive ICF \cite{lindl-nif-pop-2004}.

A wide array of computational tools is used to model LPI, ranging from
rapid ($\sim$secs) calculations of linear gains along 1D profiles to
massively-parallel kinetic particle-in-cell simulations. We present
here a new tool, called \dep, to the less computationally expensive
end of this spectrum. \dep{} solves for the pump intensity and
scattered-wave spectral density for a set of scattered frequencies, in
steady-state, along a 1D profile of plasma conditions. Pump depletion
is included, and the plasma waves are assumed to be in the strong
damping limit (i.e., they do not advect).  Fully kinetic (although
linear) formulas are used for various quantities like the coupling
coefficient. Bremsstrahlung noise and damping, as well as Thomson
scattering (TS), are included. The \dep{} model, especially the noise
sources, in some ways resembles that of Ref.\
\cite{berger-srsnoise-pofb-1989}. Other similar works which have
influenced our thinking, and use 1D coupled-mode equations, are Refs.\
\cite{ramani-sbs-pof-1983}-\cite{mounaix-lpi-pre-1997}.

\dep{} is a 1D model, but the plasma conditions are generally found by
tracing 3D geometric-optics ray paths through the output of a
radiation-hydrodynamics code. We therefore call this combined approach
to studying LPI a ray-based one. Details of this methodology, and its
limits, are discussed in Sec.\ \ref{s:ray}.

\dep{} is similar to the code \lip, which calculates linear gains for
SRS and SBS along 1D profiles (\lip{} is discussed here in Appendix A). Both codes
take seconds to analyze one profile from
the laser entrance to the high-Z wall in an ICF ignition
design. However, \dep{} includes substantially more physics than \lip,
such as pump depletion, noise sources, and
re-absorption of scattered light. \dep{} moreover provides pump
and scattered intensities, which unlike gains can be directly compared
with experiment and more sophisticated LPI codes. Despite its
simplicity, \dep{} agrees well in certain cases with
results from the 3D paraxial laser propagation code \ftd{}. This is
quite promising given \dep's much lower computing cost.

There is important physics which \dep{} does not capture, with laser
speckles or hot spots being one of the most important. Recent SBS
experiments \cite{froula-sbs-prl-2007, neumayer-sbs-prl-2008} at the
OMEGA Laser Facility \cite{boehly-omega-optcomm-1997} show good
agreement between measured reflectivity and \ftd{} predictions, while
\dep{} gives a lower value. This is due to the speckle pattern of the
phase plate smoothed lasers. Sec.\ \ref{s:nif} describes one
approximate way to bound the speckle enhancement by doubling the
coupling coefficient; the resulting \dep{} reflectivity always exceeds
the experimental level. A more sophisticated idea for handling
speckles is outlined in the conclusion. Additional beam smoothing,
like polarization smoothing (PS) and smoothing by spectral dispersion
(SSD), reduce the effective speckle intensity and can reduce the
reflectivity even below the speckle-free \dep{} level.

The paper is organized as follows. Section \ref{s:gov} derives the
governing equations for the pump intensity and scattered-wave spectral
density. Our ray-based methodology and model limits are discussed in
Sec.\ \ref{s:ray}. The numerical method is given in Sec.\ \ref{s:num},
including a quasi-analytic solution for the coupling-Thomson
step. Section \ref{s:bench} compares \dep{} with \lip{} linear gains
and \ftd{} ``plane-wave'' simulations on prescribed profiles. The
relationship between Thomson scattering and linear gain is discussed
in Sec.\ \ref{s:thom}. In Sec.\ \ref{s:omsbs} we compare \dep{} to the
experimental and \ftd{} SBS reflectivities in recent OMEGA
shots. Sec.\ \ref{s:nif} presents \dep{} analysis of an ignition
design with a 285 eV radiation temperature for the National Ignition
Facility (NIF) \cite{paisner-nif-fustech-1994}. In particular, we show
the effect of scattered light re-absorption and put a bound on speckle
enhancement.  We conclude and discuss future prospects in Sec.\
\ref{s:conc}. A review of \lip{} and its linear gain is presented in
Appendix A. Appendix B details the numerics of \dep's coupling-Thomson
step.

\section{Governing equations} \label{s:gov} 
We derive coupled-mode
equations, in time and one space dimension, for the slowly-varying
wave envelopes, and find the resulting intensity equations. We do this
for the light waves first, and then the plasma wave in the strong
damping limit. Since our approach is standard we summarize some
steps. We take these equations in steady state to apply independently
at each scattered frequency, and transition to a spectrum of scattered
light per angular frequency. This may be viewed as a ``completely
incoherent'' treatment of the scattered light at different
frequencies. Bremsstrahlung damping and fluctuations, and TS, are then
added phenomenologically. Focusing of the whole beam is finally
accounted for, giving the system \dep{} solves.  This section
culminates in the \dep{} system, Eqs.\
(\ref{eq:I0gov}-\ref{eq:i1gov}), on which some readers may wish to
focus.

\subsection{light-wave action equations}
Let $z$ be distance along the profile, and assume all wave
vectors and gradients are in $z$ ($\p_x=\p_y=0$). $z=0$ is
taken as the left edge of the domain (the ``laser entrance''), where
we specify the right-moving pump laser; we also specify boundary
values for the left-moving backscattered wave at the right edge
$z=L_z$. The light waves are linearly polarized in $y$ and represented
by their vector potentials $\vec A_i = (1/2)A_i(z,t)\hat
ye^{i\psi_i}+cc$, where $i=0,1$ for the pump and scattered wave,
respectively. $A_i$ is the slowly-varying complex envelope, and we use the
dimensionless $a_i \equiv eA_i/m_ec$. $\psi_i(z,t)$ is the
rapidly-varying phase with $k_i\equiv\p_z\psi_i$ and
$\om_i\equiv-\p_t\psi_i$. Let $\si_i\equiv k_i/|k_i|$ with
$\si_0=\si_2=+1$ and $\si_1=-1$ (appropriate for backscatter). Thermal
fluctuations give rise to both light waves and plasma
waves. However, upon appropriate averaging the \textit{field
  amplitudes} of these fluctuations vanish (but their \textit{mean
  squares} do not). The amplitudes $A_i$ (and $n_{j2}$ below)
represent only the coherent, and not the noise, components of the
fields. We insert a bremsstrahlung noise source and TS
to the intensity equations below.

From the Maxwell equations, and conservation of canonical transverse
momentum $m_ev_{ye}=eA_y$, we find $A_y=(\vec A_0+\vec A_1)\cdot\hat y$ satisfies
\begin{equation}
  \label{eq:Ay}
  \lb \p_{tt}-c^2\p_{zz}+\om_{pe}^2 \rb A_y = -\om_{pe}^2 {n_{e2}\over n_e}A_y.
\end{equation}
$\tilde n_j=n_j+N_{j2}$ is the total number density for species $j$ ($j=e$
for electrons, $i$ for an ion species),
$N_{j2}=(1/2)n_{j2}e^{i\psi_2}+cc$, and $n_{j2}$ is the slowly-varying
plasma-wave envelope. We define
$\om_{pj}\equiv[n_jZ_j^2e^2/\ep_0m_j]^{1/2}$,
$v_{Tj}\equiv[T_j/m_j]^{1/2}$ and $\la_{Dj}\equiv v_{Tj}/\om_{pj}$, with $Z_j$ the charge state. As
usual, the massive ions are treated as fixed in the transverse current. (We look
forward to a circumstance where a positively-charged species must be
considered mobile, such as an electron-positron plasma!)

Following, e.g., Ref.~\cite{dewandre-wshift-pof-1981}, we introduce
the small parameter $\de \sim \om_i^{-1} \p_t \ln X \sim k_i^{-1} \p_x
\ln X$ for $X=A_i,k_i,$ etc. We order $\p_t,\p_x \sim\de$,
$\psi_i\sim\de^{-1}$, and the right-hand side of Eq.~(\ref{eq:Ay})
$\sim\de$. To order $\de^0$, we obtain the free-wave dispersion
relation
\begin{equation}
  \om_i^2 = \om_{pe}^2 + c^2k_i^2 \qquad i=0,1.
\end{equation}
For the steady-state conditions considered below we take $\om_i$ to be
constant and find the eikonal $ck_i(x)=\si_i\eta_i\om_i$ with
$\eta_i\equiv [1-n_e/n_{ci}]^{1/2}$ and
$n_{ci}\equiv\om_i^2\ep_0m_e/e^2$ the critical density of mode
$i$. Also, the group velocity is $v_{gi}\equiv \si_i\eta_ic$.

Assuming perfect phase matching ($k_0=k_1+k_2$, $\om_0=\om_1+\om_2$),
the resonant order $\de$ terms in Eq.~(\ref{eq:Ay}) yield the envelope
equations:
\begin{eqnarray}
  \label{eq:L0a0}
  L_0 a_0 &=& -{i\over4}{\om_{pe}^2\over\om_0}{n_{e2}\over n_e}a_1, \\
  \label{eq:L1a1}
  L_0 a_1 &=& -{i\over4}{\om_{pe}^2\over\om_1}{n_{e2}^*\over n_e}a_0.
\end{eqnarray}
The operator $L_i \equiv \p_t + v_{gi}\p_z +
(1/2\om_i)(\p_t\om_i+c^2\p_zk_i)$.
Our quasi-monochromatic light waves ($i=0,1$) have action density
\cite{bers-leshouches} $N_i\equiv (m_e/8\pi r_e)\om_ia_ia_i^*$
where $r_e\equiv e^2/4\pi\ep_0m_ec^2\approx2.82$ fm. We also define the (positive) action flux
$Z_i\equiv N_i|v_{gi}|$ and intensity $I_i\equiv \om_iZ_i$. In
practical units,
\begin{equation} \label{eq:aiIi}
  |a_i|^2 = {I_i\la_i^2 \over P_{em} \eta_i}  
\end{equation}
where $\la_i\equiv 2\pi c/\om_i$ and $P_{em}\equiv (\pi/2)m_ec^3/r_e$
$\approx 1.37\times10^{18}$ W$\cdot$cm$^{-2}\cdot\mu$m$^2$. We form
Eq.~(\ref{eq:L0a0})$\times a_0^* + cc$ and Eq.~(\ref{eq:L1a1})$\times
a_1^* + cc$ to find
\begin{eqnarray}
  -\p_tN_0 - \p_zZ_0 &=& \p_tN_1 - \p_zZ_1 = J \\
  J &\equiv& -{1\over4}m_ec^2\ \im[a_0^*a_1n_{e2}].
\end{eqnarray}

\subsection{plasma-wave action equations}
We describe the plasma waves following the dielectric operator
approach of Cohen and Kaufman \cite{cohen-drivenepw-pof-1977}:
\begin{eqnarray}
  \label{eq:epn2}
  \ep(\om_2'+i\p_t,k_2-i\p_z) n_2\ &=& \npnd, \\
  \npnd &\equiv& \chi_e(\om_2',k_2){c^2k_2^2 \over 2\om_{pe}^2} n_e a_0 a_1^*.
\end{eqnarray}
The charge-density fluctuation $n_2 \equiv -n_{e2} + \sum_iZ_in_{i2}$
experiences a ponderomotive drive $\npnd$. $\om_2'\equiv\om_2-\vec
k_2\cdot\vec u$ is the Doppler-shifted plasma-wave frequency in the
frame of the plasma flow $\vec u$ ($\om_2$ is in the lab
frame). $\ep\equiv 1+\chi$ is an operator, where the time and space
derivatives reflect envelope evolution and $\chi\equiv\sum_j\chi_j$ is
the total susceptibility. $\chi_e$ in $\npnd$ is simply a
function, not an operator. $\chi_j$ is the (linear) kinetic,
collisionless susceptibility of Maxwellian species $j$:
\begin{equation}
  \chi_j \equiv -{1\over 2k_2^2\la_{Dj}^2}Z'(\ze_j);  \qquad \ze_j \equiv {\om_2'\over k_2v_{Tj}\sqrt2}.
\end{equation}
$Z(\ze) \equiv i\pi^{1/2}e^{-\ze^2}\erfc(-i\ze)$ is the plasma
dispersion function \cite{fried-zfunc-1961} and $\erfc$
is the complimentary error function \cite{absteg}. Gauss's law
relates $n_2$ and $n_{j2}$:
\begin{eqnarray}
  \label{eq:ne2n2}
  n_{e2} &=& -(1+\chi_I)n_2,  \\ 
  n_{i2} &=& -\chi_i \lp {1 \over Z_i} + {m_e \over m_i}{\ep \over \chi_e} \rp n_2 \\
  &\approx& -{\chi_i \over Z_i} n_2,
\end{eqnarray}
with $\chi_I \equiv \sum_i\chi_i$. For SRS, where the ion motion is
negligible, we usually take $1+\chi_I\rightarrow 1$ to save computing time.

Expanding $\ep$ for slow envelope variation, and retaining only
$\ep_r\equiv\re\,\ep$ in the derivatives, gives
\begin{equation}
  \lb \p_t  + v_{g2} \p_z + \nu_2 + i\de\om_2 \rb n_2 = -i {\npnd \over \dot\ep}.
\end{equation}
$\dot\ep \equiv \p\ep_r/\p\om_2'$, $\ep'\equiv \p\ep_r/\p
k_2$, $v_{g2} \equiv - \ep'/\dot\ep$ is the plasma-wave group velocity,
$\nu_2 \equiv \im[\ep]/\dot\ep$ is the damping rate, and
$\de\om_2 \equiv -\ep_r/\dot\ep$ is the phase detuning.

We now assume the plasma wave is in the strong damping limit, where
its advection is neglected: $|v_{g2}\p_zn_2| \ll
|\nu_2+i\de\om_2| |n_2|$. This implies the instability is
below its absolute threshold so that steady-state solutions are
accessible. Also going to steady-state, we find
\begin{equation} \label{eq:n2npnd} \ep(\om_2,k_2)n_2 = \npnd.
\end{equation}
Replacing $n_{e2}$ via Eqs.~(\ref{eq:ne2n2}) and (\ref{eq:n2npnd}) yields
\begin{equation} \label{eq:J2} J = \om_0\tilde\Ga_1 Z_0Z_1.
\end{equation}
The coupling coefficient $\tilde\Ga_1$ is
\begin{eqnarray}
  \tilde\Ga_1 &\equiv& \Ga_S\im\lb{\chi_e\over \ep}(1+\chi_I) \rb \label{eq:Gam1}
                = {\Ga_Sg_\Ga \over |\ep|^2}, \label{eq:Gam1res} \\
  \Ga_S &\equiv& {2\pi r_e \over m_ec^2}{1\over\om_0}{k_2^2 \over k_0|k_1|}, \\
  g_\Ga &\equiv& |1+\chi_I|^2\im\chi_e + |\chi_e|^2\im\chi_I.
\end{eqnarray}
The second form of $\tilde\Ga_1$ exhibits the resonance for $|\ep|\ll 1$. The
over-tilde on $\tilde\Ga_1$ indicates it will be modified below to
account for beam focusing.  $\tilde\Ga_1$, and thus $J$, are usually
positive. We now have a closed system for modes 0 and 1, with no independent
equation for mode 2:
\begin{eqnarray}
  \p_tN_0 + \p_zZ_0 &=& -\om_0\tilde\Ga_1 Z_0Z_1, \label{eq:NZ0} \\
  \p_tN_1 - \p_zZ_1 &=& \om_0\tilde\Ga_1 Z_0Z_1. \label{eq:NZ1}
\end{eqnarray}

\subsection{Steady-state equations for a spectrum of scattered waves}
We transition to steady state ($\p_t=0$) and work with
intensities. Since we have assumed $\p_z\om_i=0$, we multiply
Eq.\ (\ref{eq:NZ0}) by $\om_0$ and Eq.\ (\ref{eq:NZ1}) by $\om_1$ to obtain
\begin{eqnarray}
  d_zI_0  &=& -{\om_0\over\om_1}\tilde\Ga_1 I_0I_1, \label{eq:I0} \\
  -d_zI_1 &=& \tilde\Ga_1 I_0I_1. \label{eq:I1}
\end{eqnarray}
Here and elsewhere, $d_xf(x)$ denotes the ordinary derivative of a function of one variable, while $\p_xf$ denotes the partial derivative of a function of several variables.

The bremsstrahlung source and TS are expressed in terms of spectral
density $i_1(z,\om_1)$ (intensity per angular frequency). The
scattered intensity is then $I_1=\int d\om_1\, i_1$. We take Eq.\ (\ref{eq:I1}) to apply independently at each $\om_1$, and
integrate the coupling term in Eq.\ (\ref{eq:I0}), to find
\begin{eqnarray}
  d_zI_0  &=& -\int d\om_1{\om_0\over\om_1}\tilde\Ga_1 I_0i_1, \label{eq:iI0} \\
  -\p_zi_1 &=& \tilde\Ga_1 I_0i_1. \label{eq:ii1}
\end{eqnarray}
This is a totally incoherent treatment of the scattered light at different frequencies, and is unrealistic to the extent there is spectral ``leakage'' between nearby $\om_1$ intervals due to, e.g., envelope evolution.

\subsection{Bremsstrahlung source and damping}
We incorporate electron-ion inverse-bremsstrahlung light-wave damping ($\ka_0$ and
$\ka_1$) phenomenologically for modes 0 and 1, as well as
bremsstrahlung noise ($\tilde\Sigma_1$) for mode 1, to find
\begin{eqnarray}
  d_zI_0 &=& -\ka_0I_0 -\int d\om_1{\om_0\over\om_1}\tilde\Ga_1 I_0i_1, \\
  -\p_zi_1 &=& -\ka_1i_1 + \tilde\Sigma_1 +\tilde\Ga_1 I_0i_1.  \label{eq:i1brem}
\end{eqnarray}
As for $\tilde\Ga_1$, the over-tilde on $\tilde\Sigma_1$ denotes it
will be modified due to focusing.

$I_0$ and $i_1$ represent integrals over solid angles in $k$ space,
which we now specify. Absolute solid angles are needed in the noise
sources, and cannot be simply scaled away, because scattered
intensities determine pump
depletion. We follow closely Bekefi's book
\cite{bekefi-radiation-1966} in this section. We take $I_i = \Om_i
I_{i,\Om}$ for $i=0,1$ (see Secs.\ 1.6 and 1.7 of Bekefi). $I_{i,\Om}$
is the intensity per solid angle interval $d\Om$ in $k$ space, which we assume is
constant over the solid angle $\Om_i$ that participates in the
scattering. $\Om_i$ is the local (in $z$) solid angle in the plasma, which we express in terms of a cone half-angle $\ta_{p,i}$ as
\begin{equation}
  \Om_i \equiv 2\pi(1-\cos\ta_{p,i}).
\end{equation}
From Snell's law, $\ta_{p,i}$ varies with $z$ according to
\begin{equation}
 \cos\ta_{p,i} = \begin{cases} 
     0 \quad \mathrm{if}\ n_e \geq n_{ci}\cos^2\ta_v \\
     [1-\eta_i^{-2}\sin^2\ta_v]^{1/2} \quad \mathrm{otherwise}.
  \end{cases}
\end{equation}
$n_{ci}\cos^2\ta_v$ is the ``critical density'' above which we cut off backscatter ($\tilde\Ga_1=\tilde\Si_1=\ka_1=0$). $\ta_v$ is a ``vacuum'' cone angle, which we find from the solid angle
in the beam's F-cone (for simplicity we use the same solid angle for pump and scattered light).  This is reasonable if the scattering mostly
occurs in laser speckles that are near diffraction-limited. In terms
of laser optics F-number $F$,
\begin{eqnarray}
  \cos\ta_v &\equiv& \lb 1+{1\over 4F^2} \rb^{-1/2} \approx 1-{1\over8F^2}, \\
  \Om_i^v &\equiv& 2\pi(1-\cos\ta_v)\approx{\pi\over 4F^2}.
\end{eqnarray}
The approximate forms apply for $F\gg1$.

The upshot of the solid angle discussion (see especially Eq.\ (1.133)
of Bekefi) is
\begin{equation}
  \tilde\Sigma_1 = \Om_1 j(\om_1),  
\end{equation}
where $j(\om)$ is the emission coefficient, per
$d\Om$ and in one polarization (see p.\ 134 of Bekefi):
\begin{equation}
  j(\om_i) = {\eta_i \over 12\pi^3\sqrt{2\pi}} {\om_{pe}^4 \over v_{Te}} 
             {m_er_e \over c} \sumjion{n_j\over n_e}Z_j^2\ln\La_{ej}.
\end{equation}
$\ln\La_{ej}$ is sometimes called the Gaunt
factor and resembles the Coulomb logarithm, although it arises in
calculations \textit{without} ad hoc cutoffs on impact
parameter integrals (see Chap.\ 3 of Bekefi). For the case
$\om_i>\om_{pe}$, Bekefi finds $\La_{ej}=v_{Te}/(\om_i\bmin)$ where
\begin{equation}
 \bmin = \begin{cases} 
         {\ga \over 4}{\hbar \over \sqrt{m_eT_e}} &\mathrm{if}\ T_e>77Z_j^2\ \mathrm{eV}, \\
         \lp{\ga \over 2}\rp^{5/2} Z_jr_e{m_ec^2 \over T_e} &\mathrm{otherwise}.
  \end{cases}
\end{equation}
The first, high-$T_e$ case typically applies for hohlraum conditions. The
numerical pre-factors come from a detailed binary-collision
calculation, and $\ga=e^C\approx 1.781$ where $C\approx0.577$ is the
Euler-–Mascheroni constant. Our expression for $j$ does not
include the enhanced emission for $\om_i\approx\om_{pe}$ due to
collective effects \cite{dawson-emission-pof-1962}.

We find the absorption coefficient $\ka_i$ via Kirchoff's law (see Bekefi Sec.\ 2.3):
\begin{equation}
  \ka_i = {\Om_i \over \Om_i^v} {j(\om_i) \over B_v(\om_i)}.
\end{equation}
Our $\ka_i$ equals Bekefi's $\al_\om$. $B_v$ is the vacuum blackbody spectrum for one polarization, with units $dI/(d\om\, d\Om)$:
\begin{eqnarray}
  B_v(\om) &\equiv& {\hbar \over 8\pi^3c^2} {\om^3 \over e^{\hbar\omega/T_e}-1} \\
  &\approx& {\om^2T_e \over 8\pi^3c^2} \qquad \hbar\om \ll T_e.
\end{eqnarray}
$j$ given above was found for collision durations short compared to the light-wave period, which entails the Jeans limit $\hbar\om \ll T_e$. We therefore use the approximate form of $B_v$ to obtain
\begin{equation}
  \ka_i = {\sqrt2 \over 3\sqrt\pi} {\Om_i \over \Om_i^v} 
          {r_ec\eta_i\over\om_i^2} {\om_{pe}^4 \over v_{Te}^3}
          \sumjion{n_j\over n_e}Z_j^2\ln\La_{ej}.
\end{equation}

For an optically thick plasma ($\p_zi_1=0$) with no pump ($I_0=0$), we
obtain for $i_1$ from Eq.\ (\ref{eq:i1brem}) the fluctuation level
\ioot:
\begin{equation} \label{eq:ot} \ioot \equiv {\Sigma_1\over\ka_1} =
  {\Om_1^v \over f} B_v(\om_1).
\end{equation}
$f$ and $\Si_1$ are defined in Sec.\ \ref{s:focus}. We thus recover
the blackbody spectrum, required by Kirchoff's law. The factor
$\eta_1^2$ that usually appears in the blackbody spectrum in a plasma
is absent due to our treatment of solid angles.

\subsection{Thomson scattering}
Thomson scattering (TS) refers to scattering off plasma-wave
fluctuations resulting from particle discreteness
(\cite{oberman-fluct-hpp}, p.~308).  Had we retained a separate plasma
wave equation, the fluctuations would appear in it as \v{C}erenkov
emission \cite{berger-srsnoise-pofb-1989}. It is an important noise
source for backscatter, especially for SBS. We express $\De p_1$, the
TS scattered power increment per $d\om_1$ per $d\Om_1$ ($k_1$ solid
angle), within a thin slab of width $\De z$, as
\begin{eqnarray} \label{eq:Dp1}
  \De p_1 &=& {d\si \over d\om_1d\Om_1} I_0 \\
  {d\si \over d\om_1d\Om_1} &=& n_eA(z)\De z \psi r_e^2 {S\over 2\pi}.
\end{eqnarray}
$A(z)$ is the beam area, defined in Sec.\ \ref{s:focus}. $\psi \equiv 1-\sin^2\ta_s\sin^2\ta_a$ is a geometric factor. $\ta_s$ is the angle between
$\vec k_0$ and $\vec R$, the vector from source to ``observation
point''. For a beam with large $F$, $\ta_s \sim \ta_v \ll 1$. $\ta_a$ is the
angle between $\vec R$ and the pump polarization. We usually take
$\psi=1$.

The form factor $S$ (units of time) is from
Eq.~(138) of Ref.\ \cite{oberman-fluct-hpp}, valid for arbitrary
(non-Maxwellian) distributions, generalized to multiple ion species:
\begin{eqnarray}
  {|\ep|^2 \over 2\pi}S(\vec k,\om) &=& |1+\chi_I|^2F_e + |\chi_e|^2\sumjion{n_j\over n_e}Z_j^2 F_j \\
  F_j &\equiv& \int d^3v\ f_j(\vec v)\de(\om+\vec k\cdot\vec v).
\end{eqnarray}
$f_j$ is the distribution function of species $j$ $(\int d^3v\ f_j=1)$. For a Maxwellian,
\begin{equation}
  F_j={1\over kv_{Tj}\sqrt{2\pi} } e^{-\ze_j^2} = {(k\la_{Dj})^2 \over \pi\om} \im\chi_j,
\end{equation}
and
\begin{equation}
  {\om|\ep|^2 \over 2(k\la_{De})^{2}} S = g_\tau \equiv |1+\chi_I|^2\im\chi_e + |\chi_e|^2\sumjion{T_j\over T_e}\im\chi_j. 
\end{equation}
This form agrees with the multiple-ion result in Eq.~(3) of
Ref.~\cite{evans-thom-ppcf-1970}. Henceforth we assume Maxwellian distributions.

From Eq.\ (\ref{eq:Dp1}) we form a differential equation for $i_1$ that describes TS:
\begin{equation}
 \left. \p_zi_1 \right|_{TS} = \tau_1I_0 \qquad \tau_1 \equiv {\Om_1 \over AI_0} {\De p_1 \over \De z}.  
\end{equation}
Since TS transfers energy from the pump to the
scattered waves, we include it in both equations:
\begin{eqnarray}
  d_zI_0  &=& -\ka_0I_0 - \int d\om_1 {\om_0\over\om_1}I_0(\tau_1 + \tilde\Ga_1 i_1),  \label{eq:thom0} \\
  -\p_zi_1  &=& -\ka_1i_1 + \tilde\Sigma_1 + I_0(\tau_1 + \tilde\Ga_1i_1).  \label{eq:thom1}
\end{eqnarray}
For conevience we write $\tau_1$ as
\begin{eqnarray}
  \tau_1 &=& \Om_1n_er_e^2\psi {S(k_2,\om_2') \over 2\pi} = {\tau_Sg_\tau \over |\ep|^2}, \\
  \tau_S &\equiv& {\Om_1\psi \over \pi}n_er_e^2{(k_2\la_{De})^2 \over \om_2'}.
\end{eqnarray}
$\tau_1$ is always positive, while $\tau_S$ and $g_\tau$ have the same sign as $\om_2'$ (which can be negative for IAW's when the plasma flow is supersonic along $\vec k_0$).

It is useful to note that $i_\tau\equiv \tau_1/\Ga_1$ sometimes plays the role of an effective seed level for $i_1$:
\begin{equation} \label{eq:itau}
  i_\tau \equiv {\tau_1 \over \Ga_1} = {\tau_Sg_\tau \over \Ga_Sg_\Ga}.  
\end{equation}
For the special case $T_i=T_e$, we have $g_\tau=g_\Ga$ and $i_\tau$ is independent of $\chi_j$:
\begin{equation} \label{eq:itauTeTi}
  i_\tau = {\tau_S\over\Ga_S}={\Om_1\psi\over(2\pi)^3}{\om_0\over\om_2'}T_ek_0|k_1|, \qquad T_i=T_e.  
\end{equation}
This fact is used in Sec.\ \ref{s:thom} to discuss the relation of TS to linear gain.

\subsection{Whole-beam focusing} \label{s:focus}
We wish to incorporate the effects of whole-beam focusing in a simple
way. The equations as written hold locally in $z$, but do not model
focusing. To do this, we treat the transverse intensity patterns of
$I_0$ and $I_1$ to be uniform flattops of varying area $A(z)$. The
beam focuses at the focal spot $z_F$, where $A$ attains its minimum
$A(z_F)$. Let $\tilde I_i \equiv I_i(z)/f(z)$ be the total power at
$z$ divided by the focal spot area, with
focusing factor $f \equiv A(z_F)/A(z) \leq1$. We typically employ for $f$ the result for the on-axis intensity of a gaussian beam \cite{milonni-lasers-1988}:
\begin{equation}
  f = [1 + (z-z_F)^2/z_0^2]^{-1}  
\end{equation}
where $z_0$ is an effective Rayleigh range. For a Gaussian beam with optics F-number $F$, $z_0=(4/\pi)\la F^2$. This form approximately fits the random phase plate (RPP) smoothed beams designed for NIF (for an appropriate $z_0$).

Substituting $(I_0,i_1) =f \cdot(\tilde I_0,\tilde i_1)$ into Eqs.~(\ref{eq:thom0}-\ref{eq:thom1}), and freely
commuting $f$ with $\partial_z$, yields the principal equations
solved by \dep:
\begin{eqnarray}
  d_zI_0(z) &=& -\ka_0I_0 - I_0\int d\om_1\ {\om_0\over\om_1}(\tau_1 + \Ga_1i_1) \label{eq:I0gov},  \\
  \p_zi_1(z,\om_1) &=& \ka_1i_1 -\Sigma_1 - I_0(\tau_1 + \Ga_1i_1). \label{eq:i1gov}
\end{eqnarray}
$\Ga_1 \equiv f\tilde\Ga_1$ and $\Sigma_1\equiv f^{-1}\tilde\Sigma_1.$ In Eqs.~(\ref{eq:I0gov}-\ref{eq:i1gov}) and henceforth, all $I_i$
and $i_1$ are understood to have suppressed over-tildes, that is,
to refer to total transverse powers over focal-spot area. Similarly, the plasma-wave amplitude from Eq.\ (\ref{eq:n2npnd}) can be written
\begin{equation}
  {n_2 \over n_e} = {1\over2}{\chi_e\over\ep} \lb{ck_2\over \om_{pe}}\rb^2 f\ \tilde a_0 \tilde a_1^*  
\end{equation}
with $\tilde a_i^2\equiv \tilde I_i\la_i^2/(P_{em}\eta_i)$; see Eq.\ (\ref{eq:aiIi}).

All symbols in Eqs.~(\ref{eq:I0gov}-\ref{eq:i1gov}) are positive,
except $\Ga_1$ may be negative for SBS in case $\om_2'<0$. This
corresponds to the scattered wave having a higher frequency than the
pump, in the plasma frame. The scattered wave then gives energy to the
pump, and \dep{} handles this situation correctly.

\section{Ray methodology and model limits} \label{s:ray} 

\dep{} calculates LPI along given plasma conditions for a 1D profile. A typical application is to study a laser beam
propagating through conditions given by a rad-hydro simulation. We use
many independent rays to model the whole beam, which introduces some
statistical inaccuracy.  The rays are generally found by tracing
3D refracted paths through the rad-hydro output. Although strictly not a
part of \dep{}, this is the major way we utilize geometric-optics
rays. Wave-optics effects, such as laser speckles and diffraction (of
both the pump and scattered light), are also not included in \dep{}.
We present one way to approximate gain enhanacement due to speckles in
Sec.\ \ref{s:omsbs}.

However, laser intensity is \textit{not} found from a rad-hydro
simulation. Such codes generally treat a laser beam as a set of rays,
which are absorbed as they trace out refracted paths. The laser
intensity in a zone is found by dividing the total power of all rays
crossing that zone by its transverse area. This approach suffers from
several problems for our purposes, including the fact that intensities
remain finite at caustics only due to the finite number of rays and
zone size. Instead, we run \dep{} separately for each ray, and use a
model for the laser beam to give an initial intensity (at a
sufficiently low density that little absorption has occurred) and
$z$-dependent focusing factor (generally based on vacuum
propagation). The intensity along a \dep{} 1D profile is thus independent of
refraction that occurs due to the plasma. Refractive changes in beam
intensity occur, for instance, when a beam propagates between two
high-density regions. However, our independent-ray treatment has the
benefit that caustics pose no problem.

\dep{} assumes that the laser and scattered light follow the same
path, and thus see the same plasma conditions. The two light waves
refract differently if their wavelengths differ, as in SRS, or in SBS
for certain transverse plasma flows \cite{hinkel-flow-pop-1999}. The
departure of ray paths becomes significant when the two rays see
sufficiently different plasma conditions in the gain region for a
given wavelength that the coupling or other coefficients differ
significantly. This requires sufficiently strong transverse plasma
gradients.

\section{Numerical method} \label{s:num} We solve the \dep{} system Eqs.\
(\ref{eq:I0gov}-\ref{eq:i1gov}) from the laser entrance $z=0$ to the
right edge $z=L_z$. For backscatter (considered in this paper),
we give $I_{0L}$ and $i_{1R}(\om_1)$ as boundary conditions, where
$f_L\equiv f(z=0)$ and $f_R\equiv f(z=L_z)$. We solve this two-point
boundary value problem via a shooting method, marching from right to
left. We guess $I_{0R}$ and solve the initial value problem from
$z=L_z$ down to $z=0$, and iterate until the resulting $I_{0L}$ is
sufficiently close to the desired value. Because $I_{0R}$ is just one
scalar, it is more feasible to shoot on it than on the set of values
$i_{1L}(\om_1)$. Generalizing our approach to 3D, where one would have
to shoot on $I_{0R}(x,y)$ over a transverse plane, is much more
difficult; a different technique for 3D pump depletion is used in the
code \slip{} \cite{froula-lengthlim-prl-2008}. For the right-boundary seed value
$i_{1R}$, we either use 0 or the optically-thick \ioot{} from Eq.\
(\ref{eq:ot}). The choice seems to have little effect, since volume
sources (either TS or bremsstrahlung) typically
produce a comparable or larger noise level after a short distance.

We solve Eqs.~(\ref{eq:I0gov}-\ref{eq:i1gov})
by operator splitting
\cite{strang-splitting-siamjna-1968,yanencko-fracstep-1970}. Let the
operator $B$ solve the ``bremsstrahlung'' system
\begin{eqnarray}
  d_zI_0 &=& -\ka_0I_0     \label{eq:brem0}, \\
  \p_zi_1 &=& \ka_1i_1 -\Sigma_1   \label{eq:brem1},
\end{eqnarray}
and the operator $C$ solve the ``coupling-Thomson'' system
\begin{eqnarray}
  d_zI_0 &=& - I_0\int d\om_1\ {\om_0\over\om_1}(\tau_1 + \Ga_1i_1), \label{eq:coup0}\\
  \p_zi_1 &=& - I_0(\tau_1 + \Ga_1i_1) \label{eq:coup1}.
\end{eqnarray}
To advance the solution from the discrete gridpoint $z^n$ down to
$z^{n-1}$ (the decreasing index matches \dep's right-to-left
marching), we first apply $B$ for a half-step, then $C$ for a full
step, then $B$ for a half-step again. The splitting theorem guarantees
that if $B$ and $C$ are second-order accurate operators, then
the overall step is second-order accurate. Schematically, a
complete step is
\begin{equation}
  \{I_0,i_1\}^{n-1} = B_{1/2}C_1B_{1/2} \{I_0,i_1\}^n.   \label{eq:BCBstep}
\end{equation}

In usual applications we are given plasam conditions, and thus the
coefficients in the \dep{} equations, only at a discrete set of points
$\{z^n\}$. We use linear interpolation to find the coefficients at the
needed intermediate points, as shown below. We stress that the
numerical accuracy of \dep{} is strongly influenced by the quality of
the given plasma conditions.

\subsection{The bremsstrahlung step $B$}
$B$ must solve Eqs.~(\ref{eq:brem0}-\ref{eq:brem1}) with $\ka_i$ and
$\Sigma_1$ constant, to at least second-order accuracy. This linear
system is readily solved analytically. Since there are two
``half-steps'' of $B$ in Eq.\ (\ref{eq:BCBstep}), we consider a
generic step of size $\De z$ with initial conditions $\{I_0,i_1\}^1$,
yielding new values $\{I_0,i_1\}^0$ . $X^{1/2}=(X^0+X^1)/2$ denotes
the zone-centered value of some quantity $X$. If $\ka_1^{1/2}\neq0$,
we find
\begin{eqnarray}
  I_0^0 &=& I_0^1\exp[\ka_0^{1/2}\De z],  \label{eq:bremsol0} \\
  i_1^0 &=& (i_1^1-i_1^{\mathrm{OT},1/2})\exp[-\ka_1^{1/2}\De z] + i_1^{\mathrm{OT},1/2}  \label{eq:bremsol1}.
\end{eqnarray}
Eq.~(\ref{eq:bremsol1}) applies separately at each $\om_1$. For the
special case $\ka_1^{1/2}=0$, Eq.~(\ref{eq:bremsol1}) is replaced with
\begin{equation}
  i_1^0 = i_1^1 + \Sigma_1^{1/2}\De z \qquad (\ka^{1/2}=0).  \label{eq:bremsol1a}
\end{equation}

The rightmost $B$ in Eq.~(\ref{eq:BCBstep}) advances the system from
$z^n$ to $z^{n-1/2}$. Accordingly, for this step, the needed
coefficients in Eqs.~(\ref{eq:bremsol0}-\ref{eq:bremsol1a}) are
interpolated at 1/4 the way from $z^n$ to $z^{n-1}$: $X^{1/2} =
[(1/4)X^{n-1}+(3/4)X^n]$ . Similarly, the leftmost $B$ in
Eq.~(\ref{eq:BCBstep}) advances the system from $z^{n-1/2}$ to
$z^{n-1}$ and uses $X^{1/2} = [(3/4)X^{n-1}+(1/4)X^n]$. In both cases
$\De z=(z^n-z^{n-1})/2$.

\subsection{The coupling-Thomson step $C$}
We now turn to the $C$ operator. $I_0$ is evolved via a
conservation law of the $C$ system,
Eqs.\ (\ref{eq:coup0}-\ref{eq:coup1}):
\begin{equation}
  d_z\lb I_0 - \int d\om_1\ {\om_0\over\om_1}i_1 \rb=0.   
\end{equation}
On the discrete $z$ grid, this gives
\begin{equation}
  I_0^{n-1} = I_0^n + \int d\om_1\ {\om_0\over\om_1}(i_1^{n-1}-i_1^n).
\end{equation}
Before doing this, we must advance $i_1$ using Eq.~(\ref{eq:coup1}) with constant
$I_0=I_0^n$ (that is, we neglect pump depletion within a zone).  This
gives rise to a numerical challenge. Namely, the coefficients $\tau_1$
and $\Ga_1$ are both proportional to $|\ep|^{-2}$, and contain a
narrow resonance where $\re\,\ep=0$ if $\im\,\ep$ is small (that is,
where the beating of the light waves drives a natural plasma
wave). Integrating through these sharp peaks with a standard ODE
method like Runge-Kutta performs very poorly unless the resonance is
well-resolved by the $z$ grid (which it usually is not). To alleviate this problem, the key
observation is that $\ep$ itself varies slowly in space, even though
$|\ep|^{-2}$ varies rapidly near resonance. We can therefore represent
$\ep$ as linearly varying with $z$ across a cell, and analytically
solve the resulting system. We merely quote the result here, and refer
the reader to Appendix B for the derivation and definition of the
relevant quantities:
\begin{equation} \label{eq:i1CTsol} i_1^{n-1} = (i_1^n+i_\tau)e^{B_\Ga
    \De w_n}-i_\tau.
\end{equation}

\section{Benchmark on linear profiles} \label{s:bench} This section
compares the results of \dep{} with those of \lip{} and \ftd{} on two
contrived profiles with weak linear gradients, one for SRS and another
for SBS. \dep{} and \ftd{} embody quite different physical models,
each with their own approximations and limitations. One can view their
favorable comparison here as a ``cross-validation'' of these models in
a regime where they should agree.

To compare with the \lip{} linear gain $G_l$ (see Appendix A), we need
a noise level against which to compare the \dep{} scattered spectrum
at the laser entrance, $i_{1L}$. For this noise level we choose
$i_1^{br}$ at $z=0$, given by solving Eq.~(\ref{eq:i1gov}) with just
the bremsstrahlung terms ($I_0\rightarrow0$):
\begin{equation}
  \p_zi_1^{br} = \ka_1i_1^{br}-\Sigma_1.  
\end{equation}
This is exactly Eq.~(\ref{eq:brem1}). We then introduce the
\dep{} gain $G_d$:
\begin{equation}
  G_d \equiv \ln {i_{1L} \over i_{1L}^{br}} = {\mathrm{``scattering''} \over \mathrm{``noise''}},
\end{equation}
where $i_{1L}$ is the solution to the full \dep{} equations. $G_l$ and
$G_d$ are exactly equal under the following conditions: there is no
pump depletion, no TS ($\tau_1=0$), no absorption of
scattered light ($\ka_1=0$), and no volume bremsstrahlung noise
($\Sigma_1=0$); the only seeding in \dep{} is then via the boundary
values $i_{1R}(\om_1)$.

\subsection{SRS benchmark} \label{s:benchsrs}
\begin{figure}
  \includegraphics[width=2.75in]{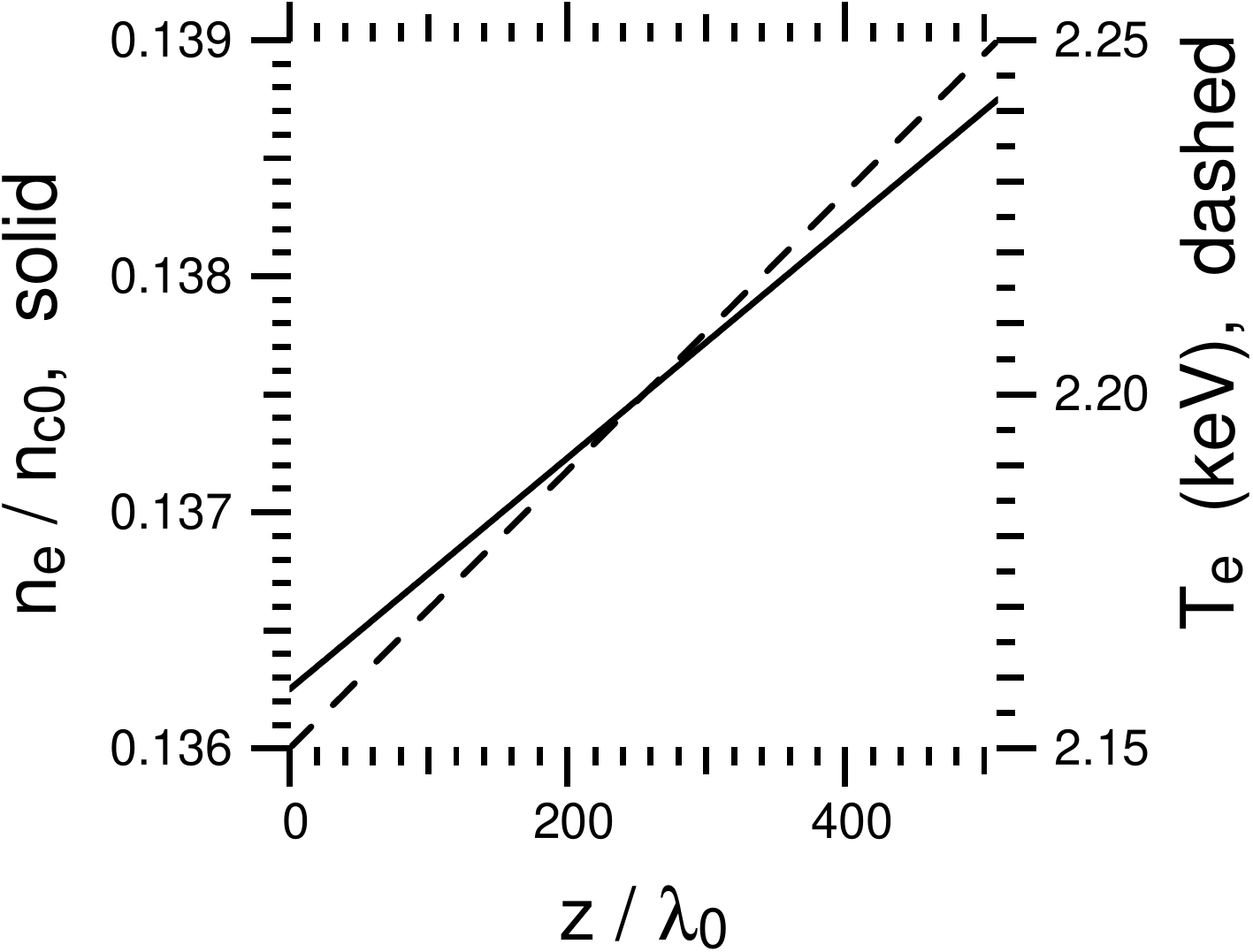}
  \caption{Plasma conditions for SRS benchmark.}
  \label{f:srsprof}
\end{figure}

\begin{figure}
  \includegraphics[width=2.3in]{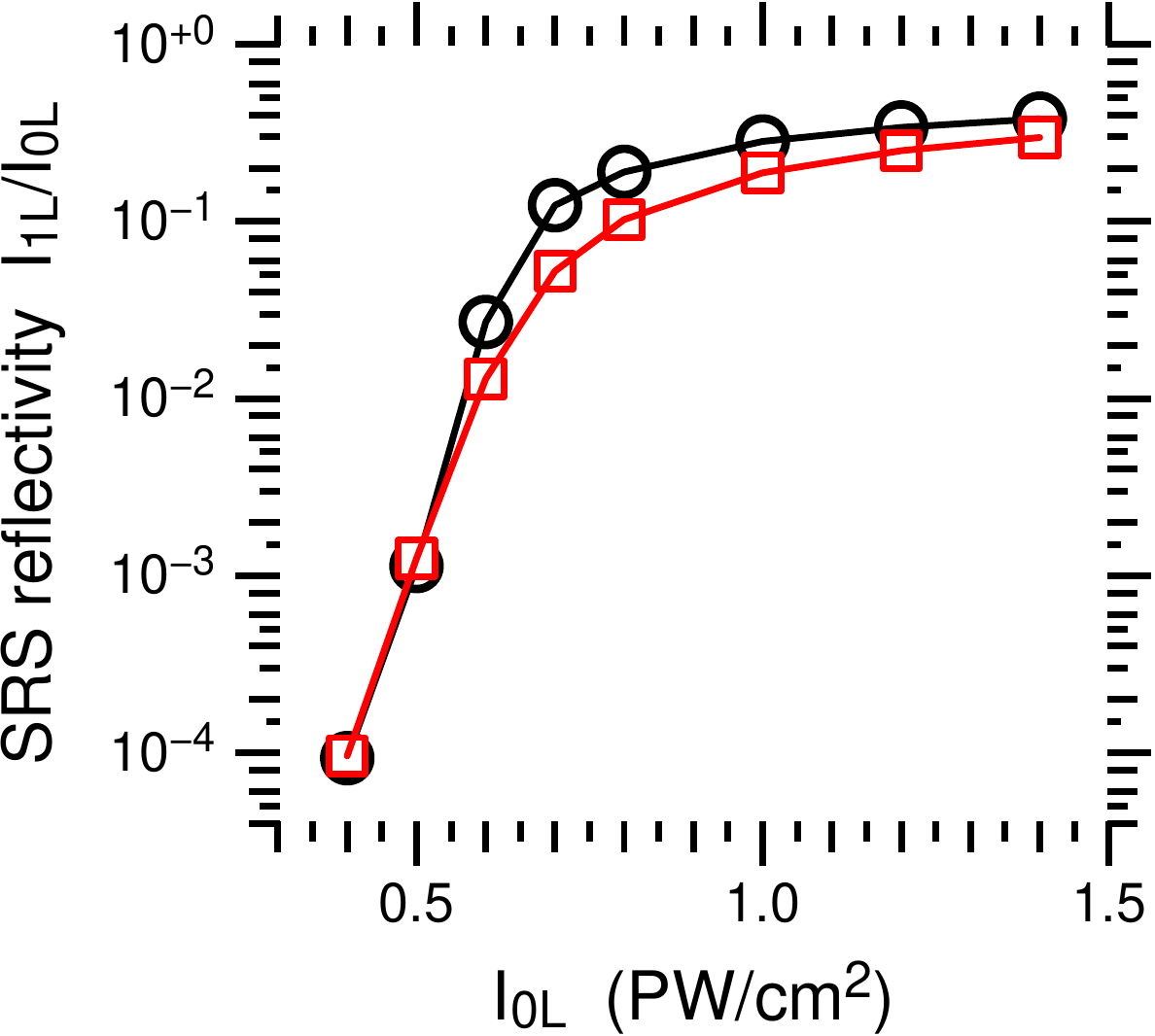}
  \caption{\colonl{} SRS reflectivity vs.\ pump intensity for the SRS benchmark
    profile of Fig.\ \ref{f:srsprof}. The black circles and red
    squares are for \ftd{} and \dep{}, respectively.}
  \label{f:srsref}
\end{figure}

The spatial profiles of our SRS benchmark plasma conditions are shown
in Fig.\ \ref{f:srsprof}. We use a profile length $L_z=510\la_0$, pump
vacuum wavelength $\la_0=(1054/3)$ nm, fully-ionized H ions with
$T_i=$ 1 keV, and no plasma flow ($\vec u=0$). In both the \dep{} and
\ftd{} runs of this section, SBS was not included. Fig.\
\ref{f:srsref} plots the resulting reflectivities for several pump
strengths. Although these are all above the homogeneous absolute instability threshold
of $I_0^{ab}\approx 0.21$ PW/cm$^2$, the time-dependent \ftd{} runs rapidly approach a steady state and show no signs of a temporally-growing mode \footnote{The homogeneous absolute instability threshold $I_0^{ab}$ is such that the undamped amplitude growth rate $\ga_0(I_0^{ab})$ satisfies $\ga_0=(1/4)|v_{g1}v_{g2}|^{1/2}(\ka_1+\ka_2)$ where $\ka_2\equiv2\nu_2/v_{g2}$ is the plasma-wave spatial energy damping rate.}. The weak gradients, or incoherent noise source, may lead to stabilization. After increasing exponentially with $I_{0L}$
for weak pumps, the reflectivity rolls over. This saturation due to
pump depletion is generic for three-wave interactions in the strong
damping limit, as demonstrated analytically by Tang
\cite{tang-sbs-jap-1966}.

\begin{figure}
  \includegraphics[width=3.25in]{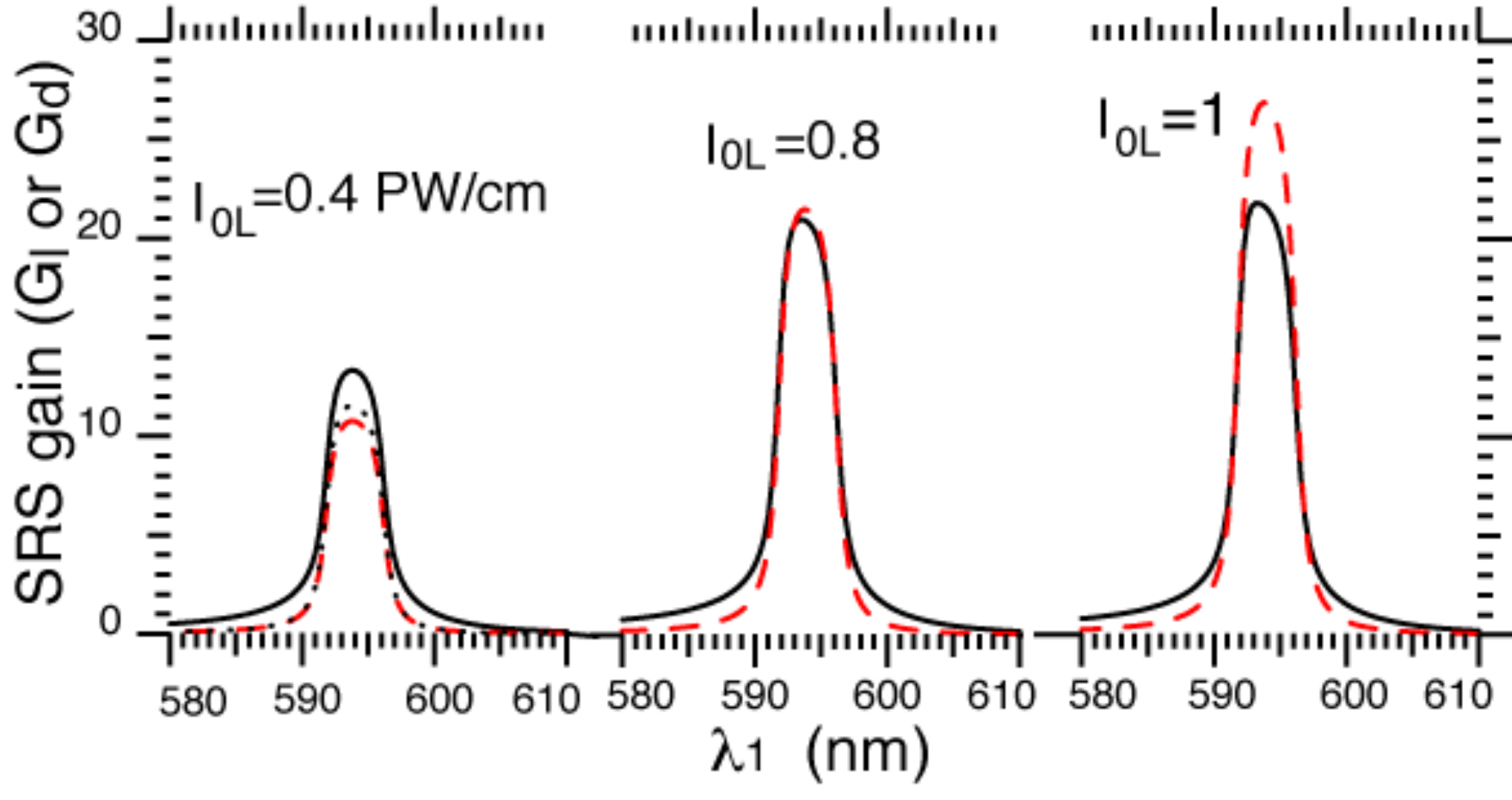}
  \caption{\colonl{} \dep{} gain $G_d$ (black solid), \lip{} gain $G_l$ (red
    dashed), and $G_d$ with no TS for $I_{0L}=0.4$
    PW/cm$^2$ ($\tau_1=0$, black dots), for SRS benchmark. TS and volume bremsstrahlung noise
    enhance $G_d$ over $G_l$ for the smallest $I_{0L}$, while pump
    depletion suppresses $G_d$ for the larger two.}
  \label{f:srsG}
\end{figure}

We compare the gains $G_l$ and $G_d$ from \lip{} and \dep{}, for
several pump strengths, in Fig.\ \ref{f:srsG}. The general shapes of
the gains are quite close, although their absolute levels differ. For
the weakest pump strength, where pump depletion plays little role (as
can be inferred from the reflectivity plot in Fig.\ \ref{f:srsref}),
the peak $G_d$ is slightly higher than $G_l$. This is due to the
volume sources in \dep{}, namely TS and bremsstrahlung
noise. To illustrate this, we plot $G_d$ found with no Thomson
scattering ($\tau_1=0$) as the black dotted curve. It lies 
between the two other curves near the peak, and overlaps $G_l$ away
from the peak. The curves for the two larger values of $I_{0L}$ in
Fig.\ \ref{f:srsG} show $G_d$ to be progressively farther below
$G_l$ at peak. This results from pump depletion, which the
reflectivity plot clearly shows is significant for $I_{0L} \gtrsim
0.8$ PW/cm$^2$. The bremsstrahlung noise level $i_1^{br}$ varies between (2.4-4.1)$\times10^{-9}$ W/cm$^2$/(rad/sec) over $\la_1=$ 650 to 550 nm.

We also compared \dep{} to the massively-parallel, paraxial laser
propagation code \ftd{} \cite{berger-f3d-pop-1998}. This code solves
for the slowly-varying envelopes of the pump laser, nearly-backscattered SRS and SBS light waves, and the daughter plasma waves,
in space and time. A carrier $\om^{en}$ is chosen for each mode
(except for the ion acoustic wave), and the corresponding rapid time
variations are averaged over. A local eikonal $k^{en}$, given by the
appropriate $\om^{en}$ and dispersion relation with local plasma
conditions, contains the rapid space variation. Kinetic quantities,
such as Landau damping rates and Thomson cross-sections, are variously
found from (linear) kinetic formulas or fluid approximations. There is
no bremsstrahlung source, but the pump and scattered light waves all
experience inverse-bremsstrahlung damping. The plasma waves undergo
Landau damping, and the advection term $v_{g2}\p_xn_2$ is retained
(i.e., they are not treated in the strong damping limit). The noise
source in \ftd{} is plasma-wave fluctuations chosen to produce the
correct TS level, and uniformly distributed over a
square in $k_\perp$ space (corresponding to the transverse $x$ and $y$
directions) extending to half the Nyquist $k$ in both $k_x$ and $k_y$.

To replicate the 1D model of \dep{}, we performed ``plane-wave''
simulations in \ftd{}. The incident laser at the $z=0$ entrance plane
is uniform in the $x$ and $y$ directions (i.e., there is no structure
like speckles), both of which are periodic with size
$L_x=L_y=128\la_0$ and grid spacing $dx=dy=1.33\la_0$. The $z$ spacing
is $dz=2\la_0$. As described above, the TS noise fills
a square in $k_\perp$ space extending to $k_x,k_y=\pm k_{1n}$, with
$k_{1n}=(3/16)k_{0v}$ and $k_{0v}\equiv\om_0/c$. We enveloped the SRS
backscattered light around $\om_1^{en}=0.592\om_0$ ($\la_1$=593.3 nm),
which has the highest linear gain. Over the slight variation of our
profile, the average $k_1^{en}=0.461k_{0v}$.

\dep{} requires a solid angle $\Om_c$, which we express in terms of an
F-number $F$, for TS and bremsstrahlung emission (we excluded the
latter for \ftd{} comparisons). Taking $k_1^{en}$ and $k_{1n}$ to
determine the focal length and spot radius, one finds
$F=k_1^{en}/2k_{1n}=1.23$. The scattered light does not uniformly fill
the noise square in $k_\perp$ space, but rather develops into a
somewhat hollow ``ring'' with a radius $\approx0.12k_{0v}$ (departing
more from a square for stronger pumps); there is some ambiguity in the
appropriate $F$ to use. We choose $F=1$, which leads to very close
reflectivities for the weakest-pump case shown in Fig.\
\ref{f:srsref}, and is near the noise-square estimate
$F=1.23$. Sidescatter at these angles may stress the accuracy of
\ftd{}'s paraxial approximation.

Figure \ref{f:srsref} shows the \dep{} and \ftd{} SRS reflectivities
for the benchmark profile. The \ftd{} values are taken at $t=$39.4 ps,
after which time all reflectivities remain roughly constant (the laser
ramped from zero to full strength over 10 ps). The agreement is quite
good, especially in the linear (weak pump) and the strongly-depleted
(strong pump) regimes. This increases confidence in the validity of
the different approximations made in both codes. It took about
2 secs of wall time for \dep{} to run on one Itanium CPU,
as opposed to 5300 secs on 16 of these CPUs for \ftd{} to advance 10 ps.

\subsection{SBS benchmark}

\begin{figure}
  \includegraphics[width=3.1in]{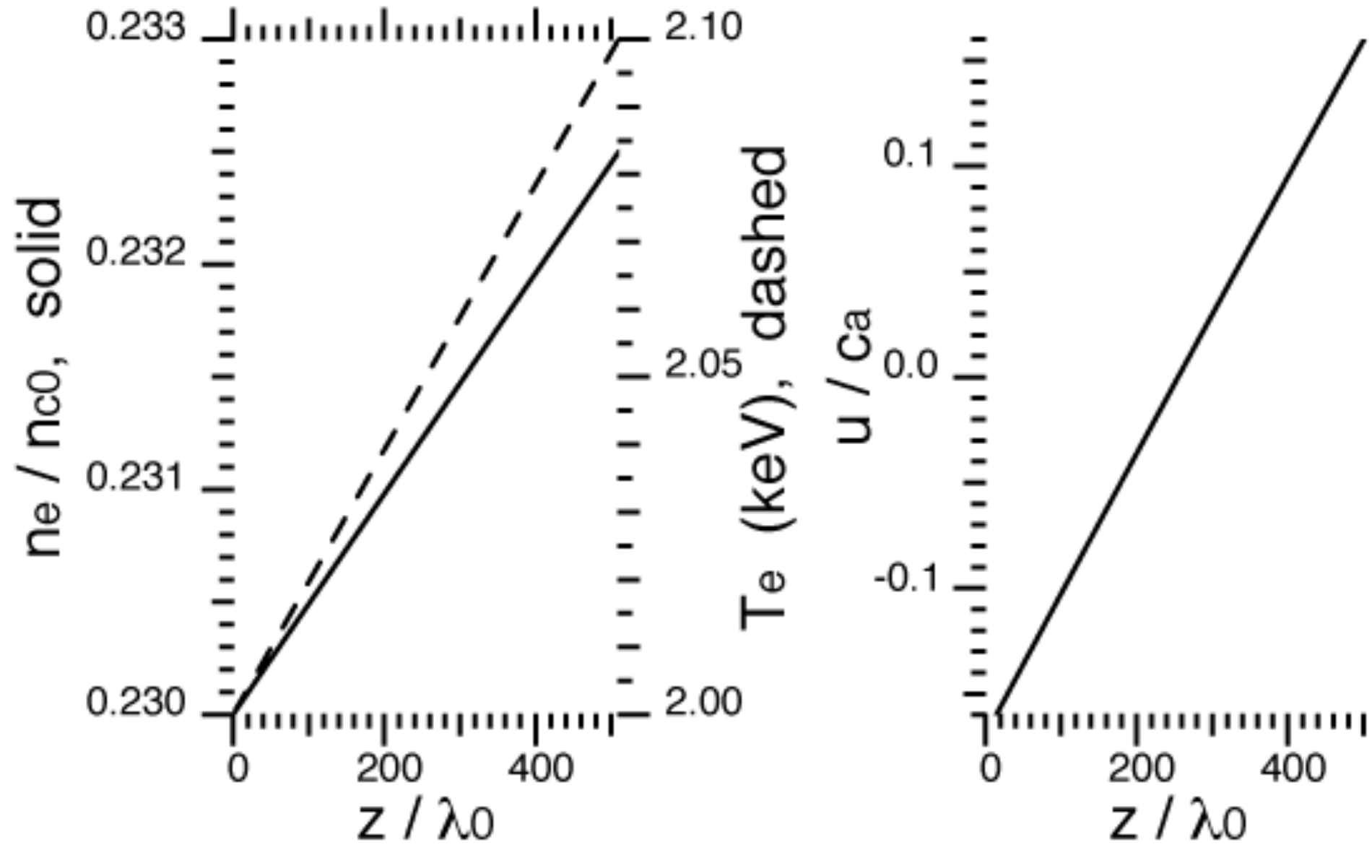}
  \caption{SBS benchmark profile.}
  \label{f:sbsprof}
\end{figure}

\begin{figure}
  \includegraphics[width=2.5in]{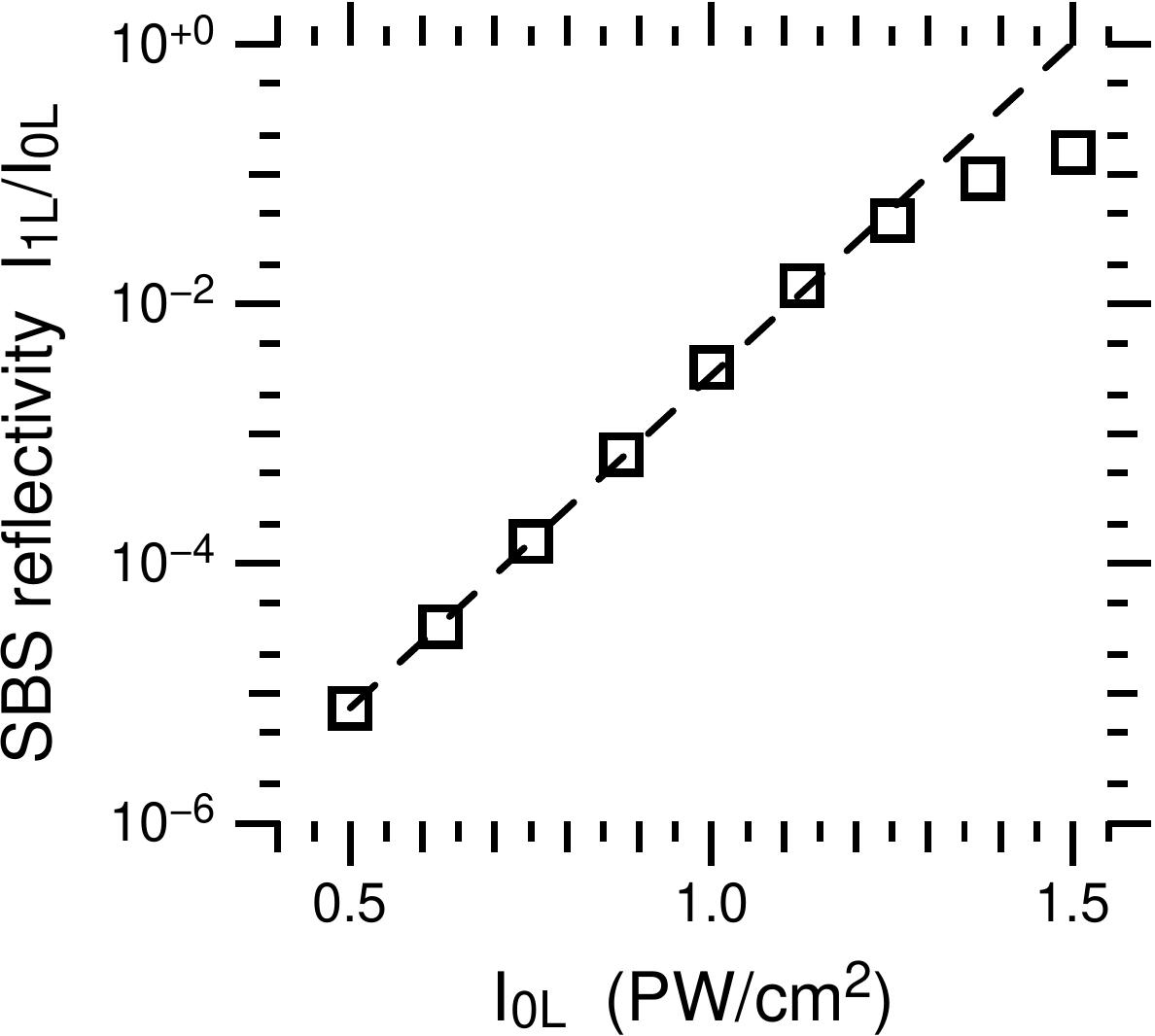}
  \caption{SBS reflectivity for SBS benchmark profile.  The
    squares are \dep{} results, and the dashed line is an extension of
    the low-$I_{0L}$ results.}
  \label{f:sbsref}
\end{figure}

We performed an SBS benchmark (with SRS neglected) using the profiles
in Fig.\ \ref{f:sbsprof}. The ions were fully-ionized He ($Z=2$,
$A=4$) with $T_i=T_e/5$. The parallel flow velocity $u$ is shown
normalized to the local acoustic speed
$c_a^2\equiv(ZT_e+3T_i)/Am_p$. The pump wavelength and profile length
match the SRS benchmark. The SBS reflectivity vs.\ pump strength is
plotted in Fig.\ \ref{f:sbsref}, which shows pump depletion for
$I_{0L}\gtrsim 1.25$ PW/cm$^2$.  We estimate the absolute threshold
$I_0^{ab}=2.6$ PW/cm$^2$ and stay below this. We used $F=1.7$ since
this gives good agreement with \ftd{} ``plane-wave'' simulations for
low $I_0$. However, for larger values of $I_0$ a ring in $k_\perp$
space develops, similar to the SRS runs, and is accompanied by a large
increase in reflectivity.

\begin{figure}
  \includegraphics[width=3.23in]{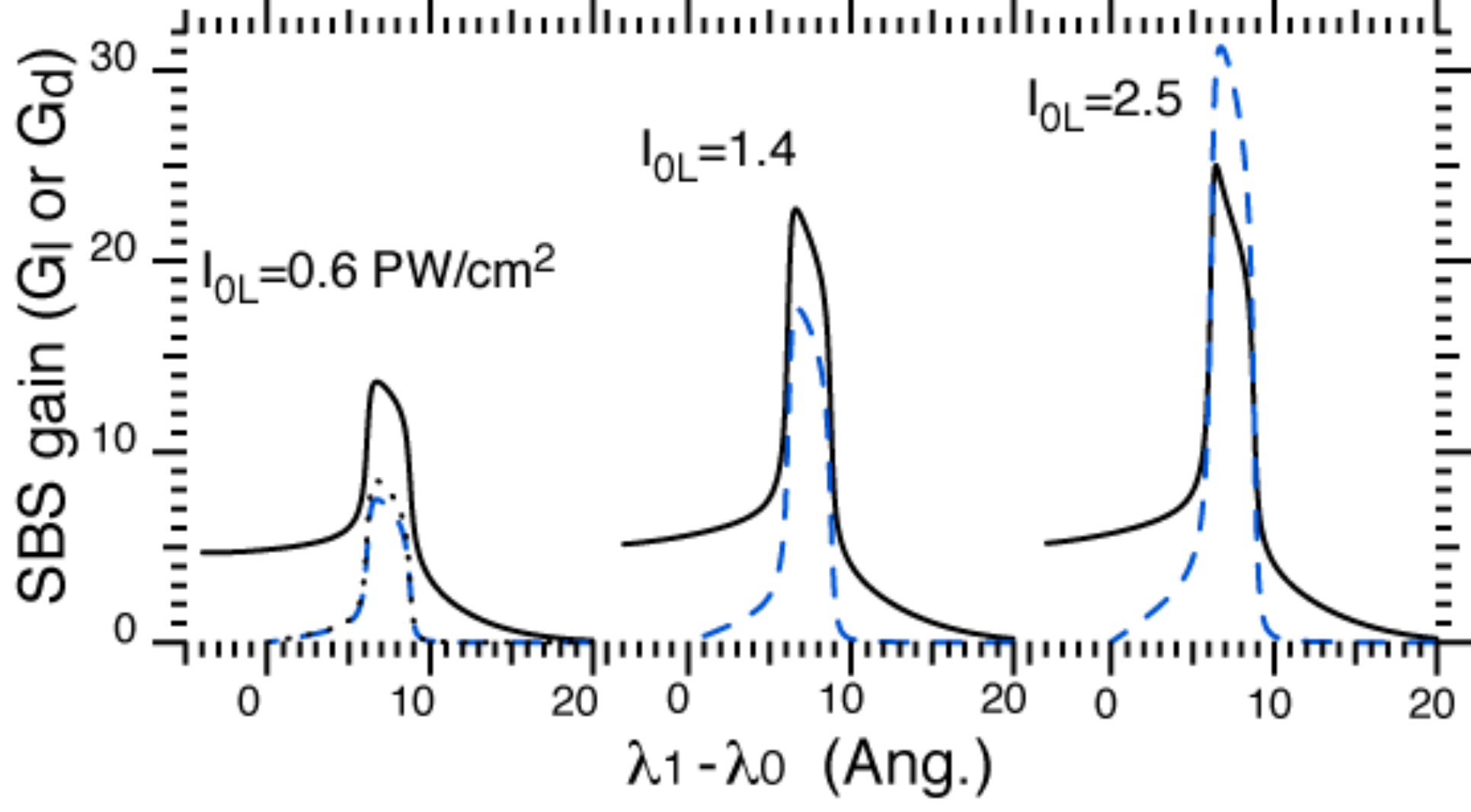}
  \caption{\colonl{} SBS \dep{} gain $G_d$ (black solid), \lip{} gain 
     $G_l$ (blue dashed), and $G_d$ without TS for 
     $I_{0L}=$0.6 PW/cm$^2$ ($\tau_1=0$, black dotted), for SBS benchmark profile.}
  \label{f:sbsG}
\end{figure}

Figure \ref{f:sbsG} compares the \dep{} and \lip{} gains, $G_d$ and
$G_l$. For the smaller two pumps we see the enhancement of $G_d$ over
$G_l$ due to TS (even though pump depletion has set in for the second case $I_{0L}=$
1.4 PW/cm$^2$), as discussed in Sec.\ \ref{s:thom}. The dotted black
curve for $I_{0L}=$ 0.6 PW/cm$^2$ is $G_d$ computed with no TS, and shows the modest
increase in $G_d$ stemming from bremsstrahlung volume (as opposed to
boundary) noise. The elevated plateau of $G_d$ to the left of the peak
is also due to TS. $I_{0L}=$2.5 PW/cm$^2$ gives $G_d<G_l$ due to
strong pump depletion. In all cases the wavelength and width of the
main peak of the two spectra are similar. $i_1^{br}$, the
bremsstrahlung solution, varies slightly from
(4.17-4.25)$\times10^{-9}$ W/cm$^2$/(rad/sec) over $\la_1-\la_0=$ 20
to -3 \AA.

\section{The relation of Thomson scattering to linear gain}  \label{s:thom}
As seen in our benchmark runs, TS leads to an enhancement of the \dep{} gain compared to the \lip{} gain (for negligible pump depletion). This is readily seen via the scattered-wave equation with just coupling and TS, Eq.\ (\ref{eq:coup1}):
\begin{equation}
      \p_zi_1 = - I_0(\tau_1 + \Ga_1i_1).
\end{equation}
We use Eq.\ (\ref{eq:itau}) to obtain
\begin{equation}
      \p_zi_1 = - \ga(i_\tau + i_1).
\end{equation}
$\ga \equiv I_0\Ga_1$ is the spatial gain rate. Typically, $\ga$ has a narrow peak in $z$ at the resonance point, while $i_\tau$ varies slowly. For simplicity, we hold $i_\tau$ constant at the resonance point, and solve for $i_1$ across the region $z=0$ to $L_z$ which includes the resonance. In our usual notation,
\begin{equation}
  i_{1L} = (i_{1R}+i_\tau)e^{G_l}-i_\tau.  
\end{equation}
$G_l\equiv \int_0^{L_z} dz\, \ga$ is the \lip{} linear gain. For $G_l\ll 1$, $i_{1L}=i_{1R}(1+G_l)+i_\tau G_l$, and emission due to the boundary source dominates over TS. In the opposite limit, 
\begin{equation}
  i_{1L} = (i_{1R}+i_\tau)e^{G_l}, \qquad e^{G_l}\gg1.
\end{equation}

TS therefore gives rise to an effective boundary
source $i_\tau$ (for a narrow resonance). In this sense, it does not
significantly alter the shape of the gain spectrum ($i_\tau$ varies
slowly with $\om_1$). However, it \textit{does} lead to a difference
in the absolute magnitude of the scattered spectrum, as embodied in
an ``absolutely-calibrated'' gain like $G_d$. As an
illustration, let us take $i_{1R}=\ioot$, the optically-thick
bremsstrahlung result of Eq.\ (\ref{eq:ot}), for simplicity evaluated at the
resonance point in the Jeans limit $\hbar\om_1\ll T_e$. Moreover, we
set $T_i=T_e$ so that $i_\tau$ assumes the simple form of Eq.\
(\ref{eq:itauTeTi}). The effective seed is then
\begin{equation} \label{eq:i1rit}
   i_{1R}+i_\tau \rightarrow \ioot\lp 1 + 
   {\Om_1\over\Om_1^v}\psi f\eta_0\eta_1{\om_0\over\om_1}{\om_0\over\om_2'} \rp.
\end{equation}
The second term on the right ($=i_\tau/\ioot$) is typically
$\lesssim10$ for SRS: for our SRS benchmark,
$i_\tau/\ioot\approx3$. But, it can be quite large for SBS since
$\om_0\gg\om_2'$ (for our SBS benchmark, $i_\tau/\ioot\approx400$). A similar result is found in Ref.\
\cite{berger-srsnoise-pofb-1989}. The authors explain this on the
thermodynamic ground that bremsstrahlung and \v{C}erenkov emission
(which produces TS) generate equal light- and plasma-wave action, so
the light-wave energy dominates by the frequency ratio. This manifests
itself in the $\om_0/\om_2'$ factor in Eq.\ (\ref{eq:i1rit}), which is much larger for SBS.

\section{Simulation of SBS experiments} \label{s:omsbs} 

Experiments have been conducted recently at the OMEGA laser to study
LPI in conditions similar to those anticipated at NIF \cite{froula-omega-pop-2007}. These shots use a gas-filled hohlraum, and a set of
``heater'' beams to pre-form the plasma environment. An
``interaction'' beam is propagated down the hohlraum axis after being
focused through a continuous phase plate (CPP) \cite{dixit-cpp-optlet-1996} with an f/6.7 lens to a
vacuum best focus of 150 $\mu$m. The plasma conditions along the
interaction beam path have been measured using TS
\cite{froula-thomson-pop-2006}, validating 2-dimensional \hydra{}
\cite{marinak_hydra_pop_2001} hydrodynamic simulations that show,
700 ps after the rise of the heater beams, a uniform 1.5-mm plasma with
an electron temperature of $\approx$2.7 keV \cite{meezan-lpi-pop-2007}.

\begin{figure}
\includegraphics[width=2.4in]{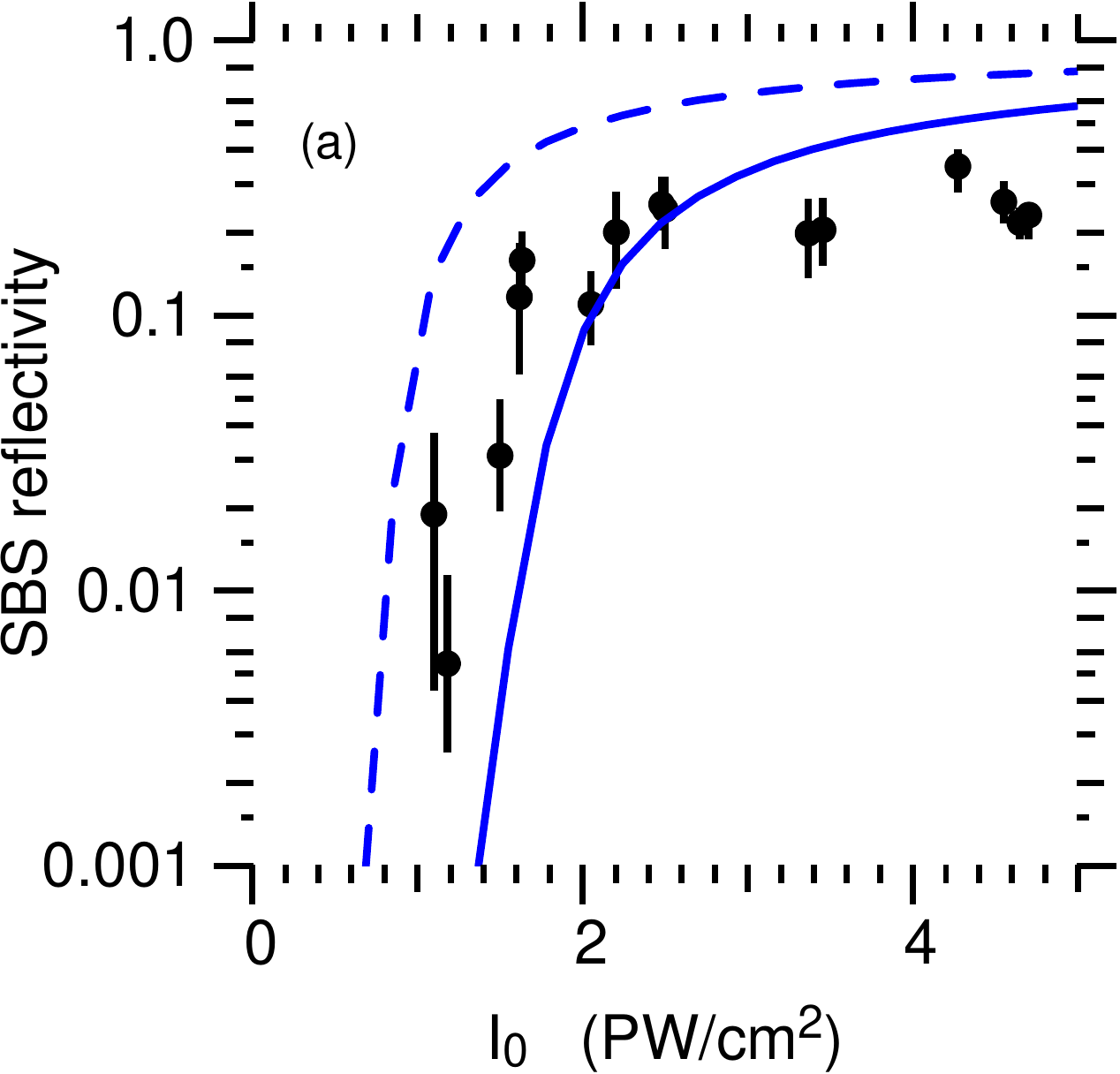}
\\ \includegraphics[width=2.4in]{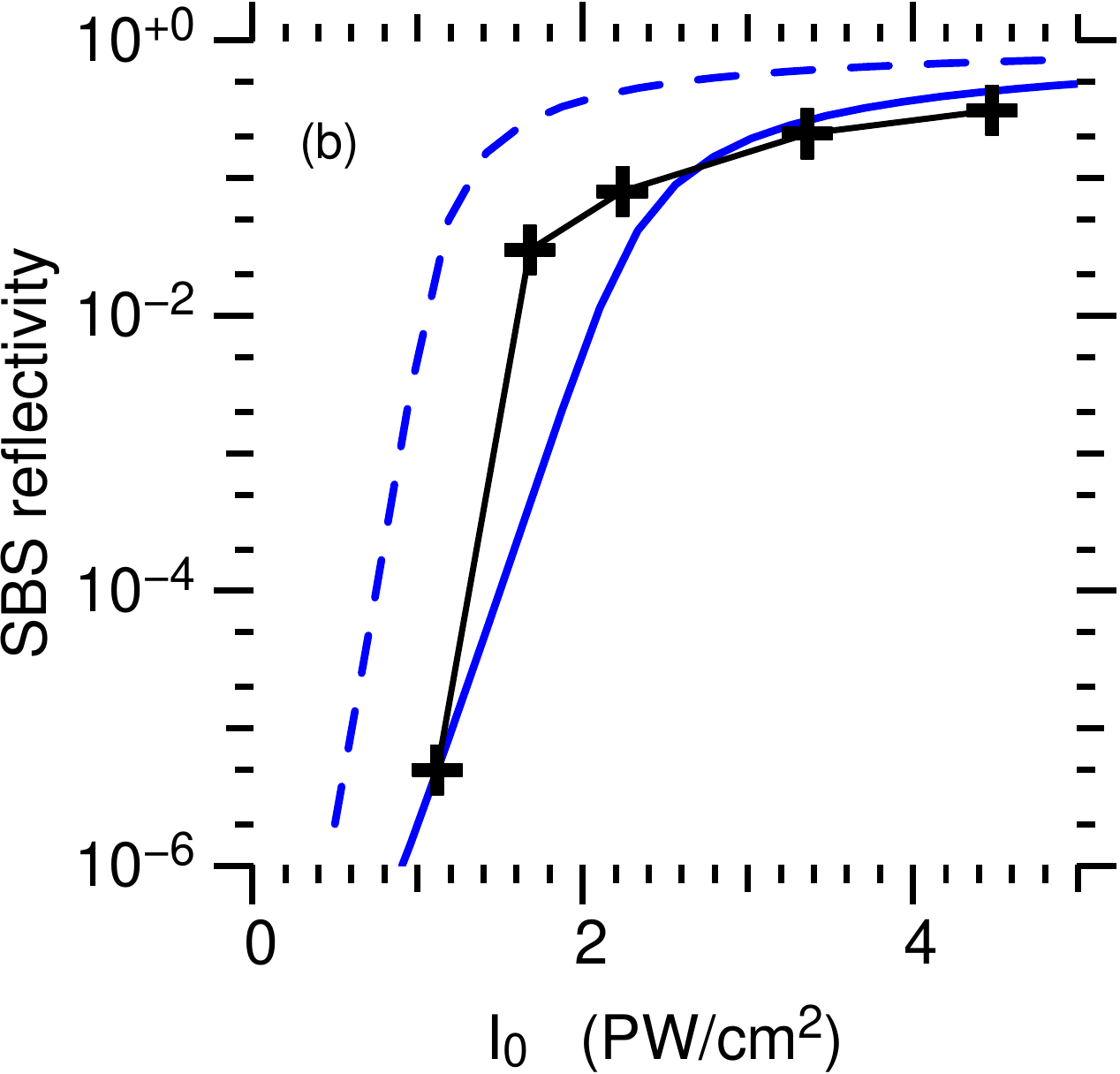}
\caption{\colonl{} (a) SBS reflectivity for OMEGA experiments with CH
  gas fill and $T_e\approx2.7$ keV (described in text). Black circles
  are measured values, the blue solid curve is \dep{} calculations
  with the nominal coupling $\Ga_1$, and the blue dashed curve is
  \dep{} calculations with $2\times\Ga_1$. (b) \dep{} and \ftd{} SBS
  reflectivities for a similar configuration but $T_e\approx3.3$
  keV. Black crosses are \ftd{} simulations, and the blue curves are
  the \dep{} results as in (a).}
\label{f:omch}
\end{figure}

Figure \ref{f:omch}(a) displays the instantaneous SBS reflectivity
increasing exponentially with the interaction beam intensity 700 ps
after the rise of the heater beams. These experiments employed a 1
atmosphere gas-fill with 30\% CH$_4$ and 70\% C$_3$H$_8$ to produce an
electron density along the interaction beam path of
0.06$n_{c0}$. Three-dimensional \ftd{} simulations agree well with the
experiments \cite{divol-aps07-pop-2008}.  Unlike the “plane-wave” simulations
discussed in Sec.\ \ref{s:benchsrs}, these simulations include the full
speckle physics. The \dep{} results (blue solid curve) fall well below
the experimental data in the regime where
pump depletion does not play a significant role ($I_0 \lesssim 2$
PW/cm$^2$). This indicates that speckles are enhancing the SBS.

One way to approximate the speckle enhancement is to consider how much
the coupling increases for the completely phase-conjugated mode
\cite{zeldovich-phconj}. This mode has a transverse intensity pattern
perfectly correlated with that of the pump, over several axial ranks
of speckles, and therefore enhances the coupling coefficient $\Ga_1$
\cite{divol-phaseconj-dpp-2005}. For an RPP-smoothed beam with
intensity distribution $\sim e^{-I/I_c}$, this effectively doubles
$\Ga_1$. This should provide an upper bound on the reflectivity so
long as the gain per speckle is $\lesssim 1$. If this is not the case,
the gain in a speckled pump suffers a mathematical divergence
(mitigated by pump depletion) as described in Ref.\ \cite{rose-div-prl-1994}. Our
phase-conjugate considerations would then not apply.

The blue dashed curve in Fig.\ \ref{f:omch} shows the \dep{}
results with twice the nominal coupling. The $2\times\Ga_1$ curve is
always above the experimental reflectivities. The threshold intensity
for which SBS equals 5\% is 1.8 PW/cm$^2$ and 0.9 PW/cm$^2$ for \dep{}
with the nominal and twice-nominal coupling, respectively, while the
experimental threshold is $\approx$1.5 PW/cm$^2$.

Comparison of \dep{} and \ftd{} is displayed in Fig.\
\ref{f:omch}(b). These calculations were performed using plasma
conditions from a \hydra{} simulation, for a configuration similar to
that of Fig.\ \ref{f:omch}(a), but with a higher heater-beam
energy. The resulting conditions are similar, except the electron
temperature is higher (about 3.3 keV). The \dep{} reflectivity with
the nominal coupling (solid blue curve) lies below the \ftd{} results
for the two intermediate values of $I_0$. This demonstrates speckle
effects enhance the \ftd{} reflectivity for moderate $I_0$. The \dep{}
results for $2\times\Ga_1$ (dashed blue curve) are always above the
\ftd{} results. Preliminary analyses with \dep{} and \ftd{} of OMEGA
experiments designed to study ion Landau damping in SBS
\cite{neumayer-sbs-prl-2008} also show a significant enhancement due
to speckles.

\section{Analysis of NIF ignition design} \label{s:nif} 
\begin{figure}
  \includegraphics[width=2.25in]{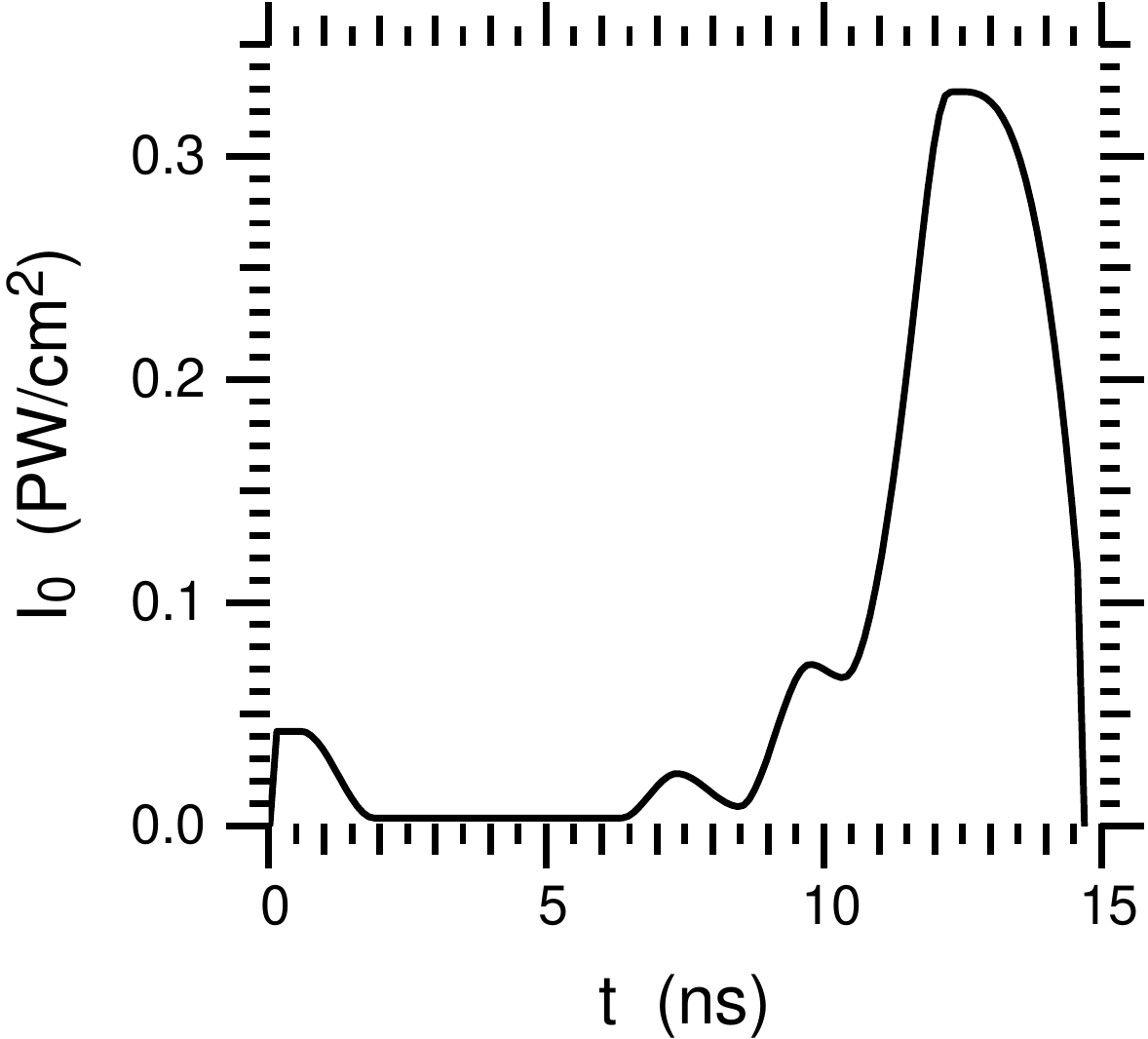}
  \caption{Nominal intensity at best focus for 285 eV NIF 
     ignition design (``NIF example''), found by dividing the laser power per quad
     by the focal spot size. The peak intensity corresponds to
     6.9 TW/quad.}
  \label{f:nifIfoc}
\end{figure}

In this
section, we exercise \dep{} on an actual NIF indirect-drive ignition
target design. The target was designed using the hydrodynamic code \las{}
\cite{zimmerman-lasnex-cppcf-1975}. For more details about the design
see Ref.\ \cite{callahan-ifsa07}; LPI analysis for this and
similar ignition targets, including massively-parallel, 3D \ftd{}
simulations, can be found in Ref.\ \cite{hinkel-aps07-pop-2008}. The
design utilizes all 192 NIF beams (at 351 nm ``blue'' light), which
deliver 1.3 MJ of laser energy. We analyze LPI along the 30$^\circ$
cone of beams (one of the two ``inner'' cones). The pulse shape for
one quad (a bundle of four beams), expressed as nominal intensity at
best focus, is shown in Fig.\ \ref{f:nifIfoc}, and reaches a maximum
of 0.33 PW/cm$^2$.  The speckle pattern for a quad approximately
corresponds to an F-number of $F=8$, which we use for \dep{}'s noise
sources (but each beam individually has $F=20$ optics). The focal spot
is elliptical with semi-axis lengths of 693, 968 $\mathrm{\mu m}$. The
peak temperature of the radiation drive is 285 eV. The materials are
as follows: the capsule ablator is Be, a plastic (CH) liner surrounds
the laser entrance hole, the hohlraum wall is Au-U with a thin outer
layer of 80\% Au-20\% B (atomic ratio), and the initial fill gas is
80\% H-20\% He. The lower-Z components are included in the last two
mixtures to reduce SBS by increasing the ion Landau damping of the
acoustic wave.

\begin{figure}
  \includegraphics[width=3in]{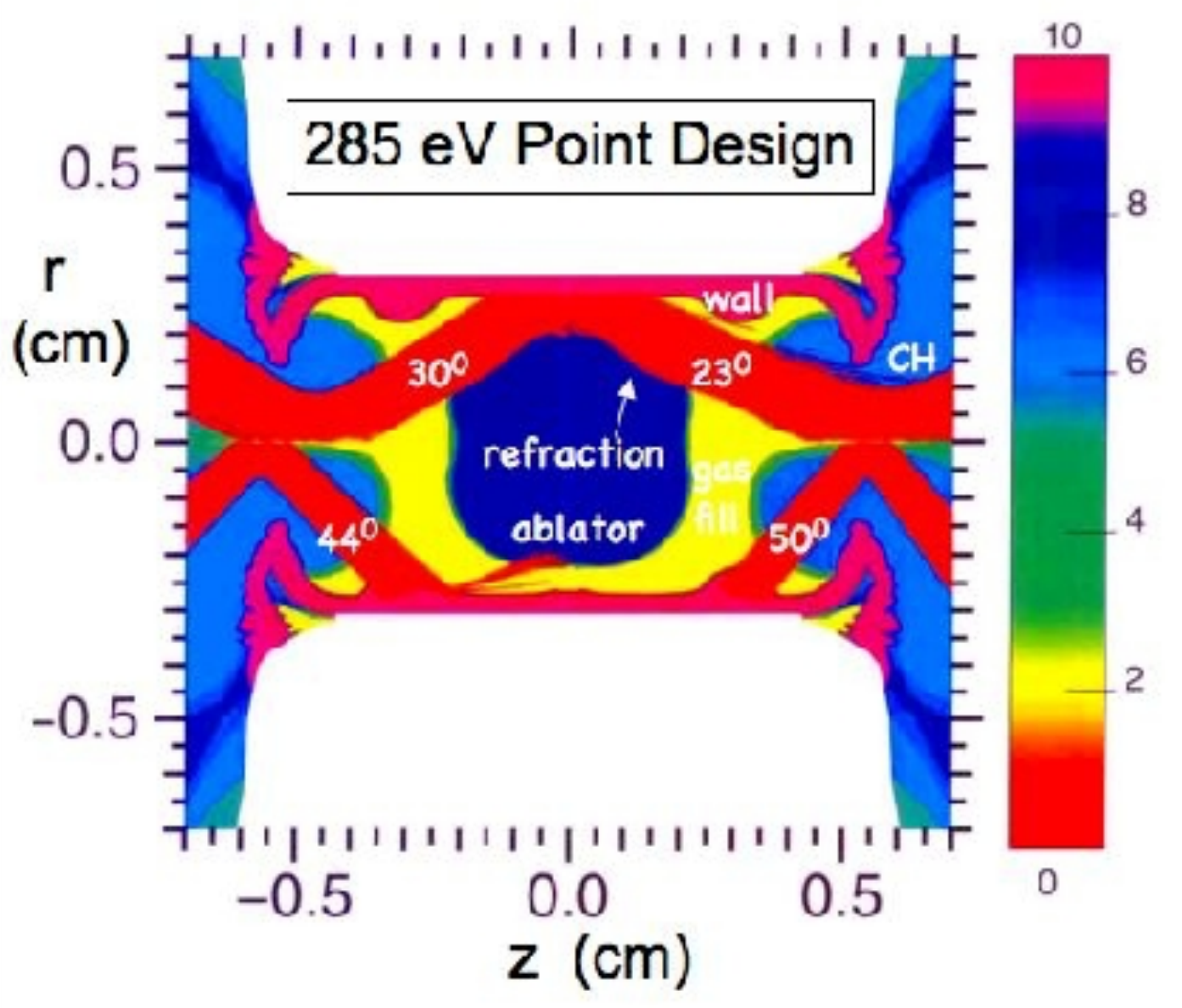}
  \caption{\colonl{} Materials and laser beam cones for NIF example.}
  \label{f:nifmat}
\end{figure}

We performed \dep{} calculations, with both SRS and SBS, at several
times and over 381 ray paths for each time. One must take an
appropriate ``average'' over the rays to characterize the LPI on a
cone. Regarding \lip{} gains, this has led to several
approaches. These include averaging the gain, finding the maximum
gain, or averaging $e^{G_l}$. This last method stems from assuming
there is no pump depletion and noise sources are independent of
scattered frequency; in this limit, the reflectivity should be roughly
proportional to $e^{G_l}$. However, this averaging, and a fortiori
taking the maximum, can be dominated by gains that are larger than
physically allowed by pump depletion or other nonlinearity. One can
attempt to include pump depletion via a Tang formula for $G_l$ at each
$\om_1$ \cite{tang-sbs-jap-1966}.

\dep{} allows for more physical ray-averaging schemes. To the extent
the transverse intensity pattern of a cone is uniform, each ray
represents the same incident laser power. Averaging \dep's ray
reflectivities then measures the fraction of incident power that gets
reflected. Pump depletion is of course included, which limits
backscatter along high-gain rays in a physical way. The reflectivities
and scattered spectra plotted here are simple averages over the rays.

\begin{figure}
  \includegraphics[width=2.4in]{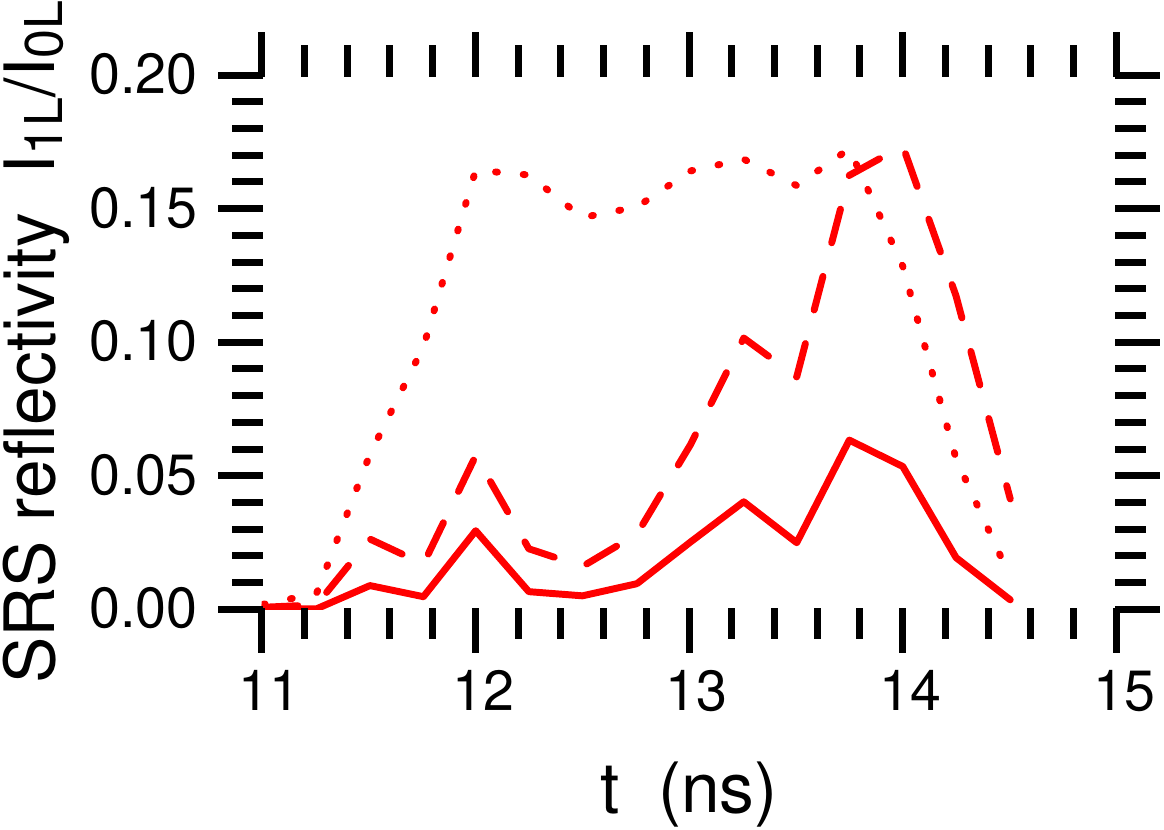} \\
  \includegraphics[width=2.4in]{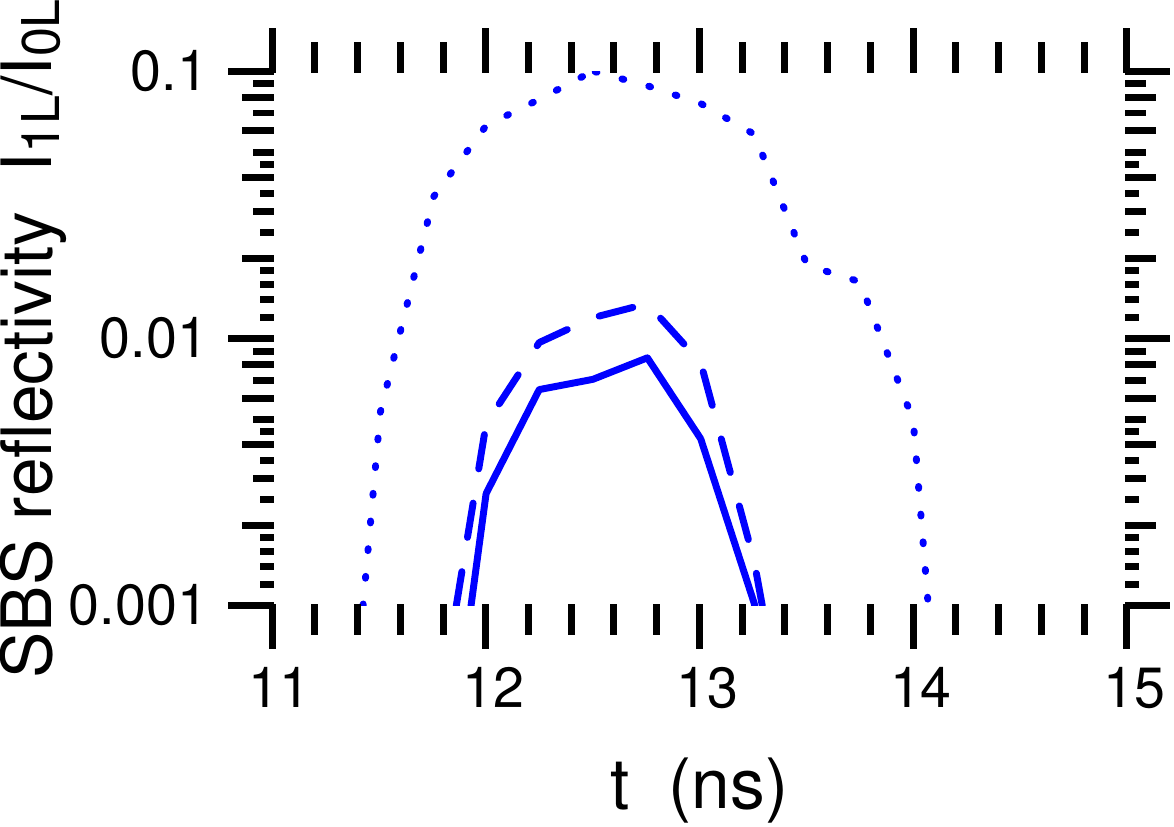}
  \caption{\colonl{} \dep{} SRS and SBS ray-averaged reflectivities
    $I_{1L}$ for NIF example.  Solid lines are the nominal case
    (re-absorption and $\Ga_1$ unscaled), dashed lines are the nominal
    $\Ga_1$ but no re-absorption of scattered light ($\ka_1=0$), and
    dotted lines are $2\times\Ga_1$ with re-absorption.}
  \label{f:nifref}
\end{figure}

\begin{figure}
  \includegraphics*[width=2.75in]{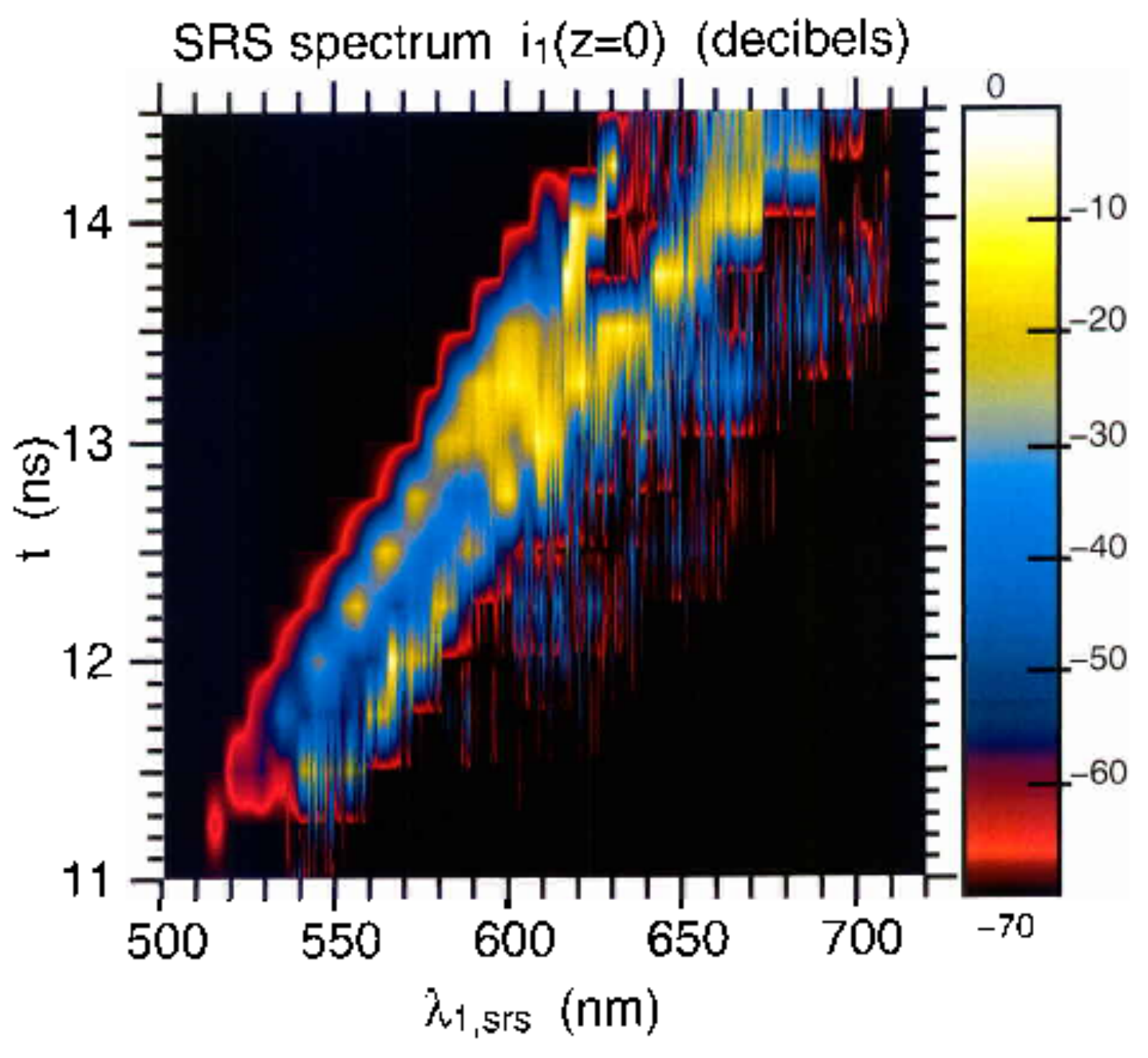}
  \caption{\colonl{} SRS streaked spectrum $i_{1L}$ for NIF example, nominal case ($\ka_1\neq0$, $1\times\Ga_1$).}
  \label{f:nifspecr}
\end{figure}

\begin{figure}
  \includegraphics*[width=2.75in]{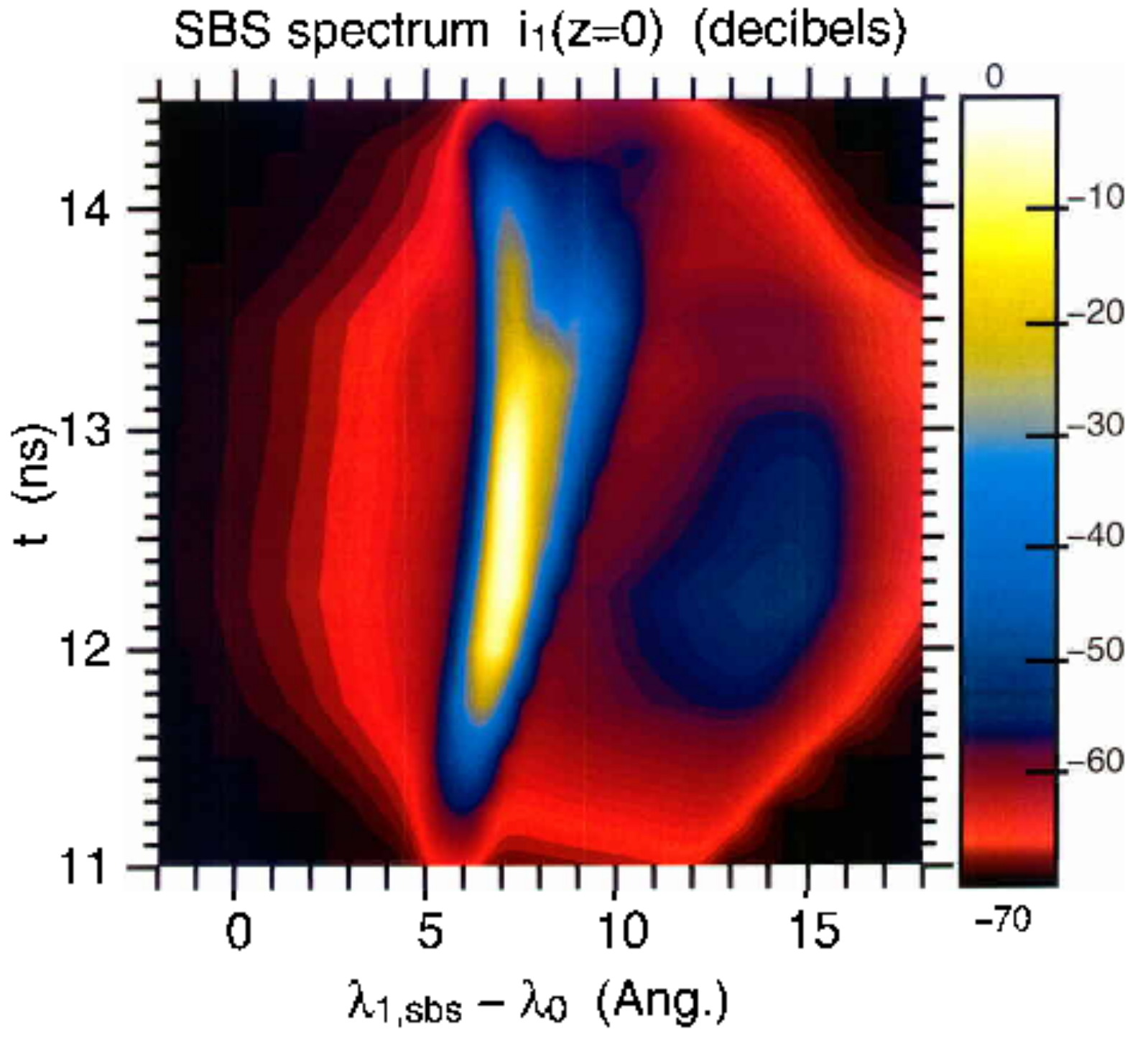}
  \caption{\colonl{} SBS streaked spectrum for NIF example, nominal case ($\ka_1\neq0$, $1\times\Ga_1$). The white-yellow streak from 5-8 \AA{} occurs in the Be ablator, while the weaker feature from 12-15 \AA{} occurs in the gas fill.}
  \label{f:nifspecb}
\end{figure}

The reflectivities for several times near peak laser
power, for the 30$^\circ$ cone, are shown in Fig.\ \ref{f:nifref}. The
results for three different cases are presented. First, the solid
lines give the reflectivities computed with the unmodified \dep{}
equations. To quantify the role of re-absorption of scattered light in
the target, we re-ran \dep{} with $\ka_1=0$. This leads to the dashed
lines. Finally, to bound the enhancement due to speckles, we plot the
results when $\Ga_1$ is doubled (and $\ka_1\neq0$) as the dotted
lines.

\begin{figure}
  \includegraphics*[width=2.75in]{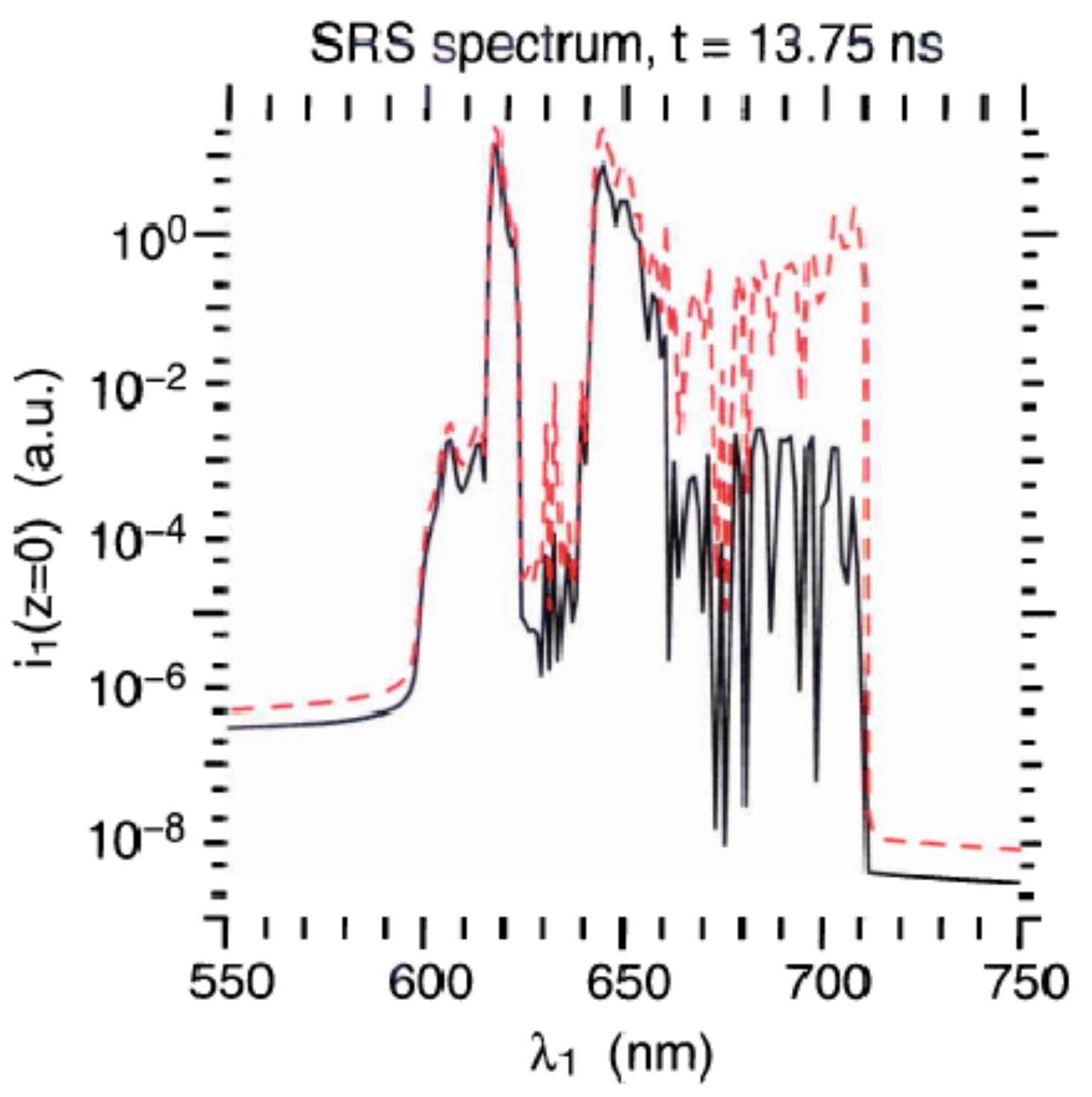}
  \caption{\colonl{} \dep{} SRS spectrum at time 13.75 ns for NIF example, smoothed over $\approx1$ nm. The black solid and red dashed lines are computed with ($\ka_1\neq0$) and without ($\ka_1=0$) re-absorption of scattered light, respectively.}
  \label{f:nifisrs}
\end{figure}

The spectra of escaping SRS and SBS light (averaged over rays) are
shown in Fig.\ \ref{f:nifspecr}-\ref{f:nifspecb}. The SBS feature at a
wavelength shift of 5-8 \AA{} comes from the Be ablator blowoff. A
much weaker feature appears from 12-13 ns at 12-15 \AA{}, and occurs
in the gas fill. The SRS spectrum is more irregular, showing two main
features separated by $\approx$20 nm that move to higher $\la_1$ as
time increases. In addition, there are narrow features at higher
$\la_1$ that originate near the hohlraum wall; these would be reduced
in a ray-averaged gain, since the exact $\la_1$ active for each ray
depends sensitively on conditions near the wall and therefore varies
from ray to ray. Re-absorption strongly suppresses these high-$\la_1$
spikes, as is seen in the SRS spectra with and without re-absorption
at $t=$ 13.75 ns in Fig.\ \ref{f:nifisrs}. Collisional plasma-wave
damping, currently not in \dep{}, may reduce the high-$\la_1$
scattering (the Landau damping of the low-$k_2\la_{De}$ plasma waves
is negligible).

\begin{figure}
  \includegraphics[width=2.4in]{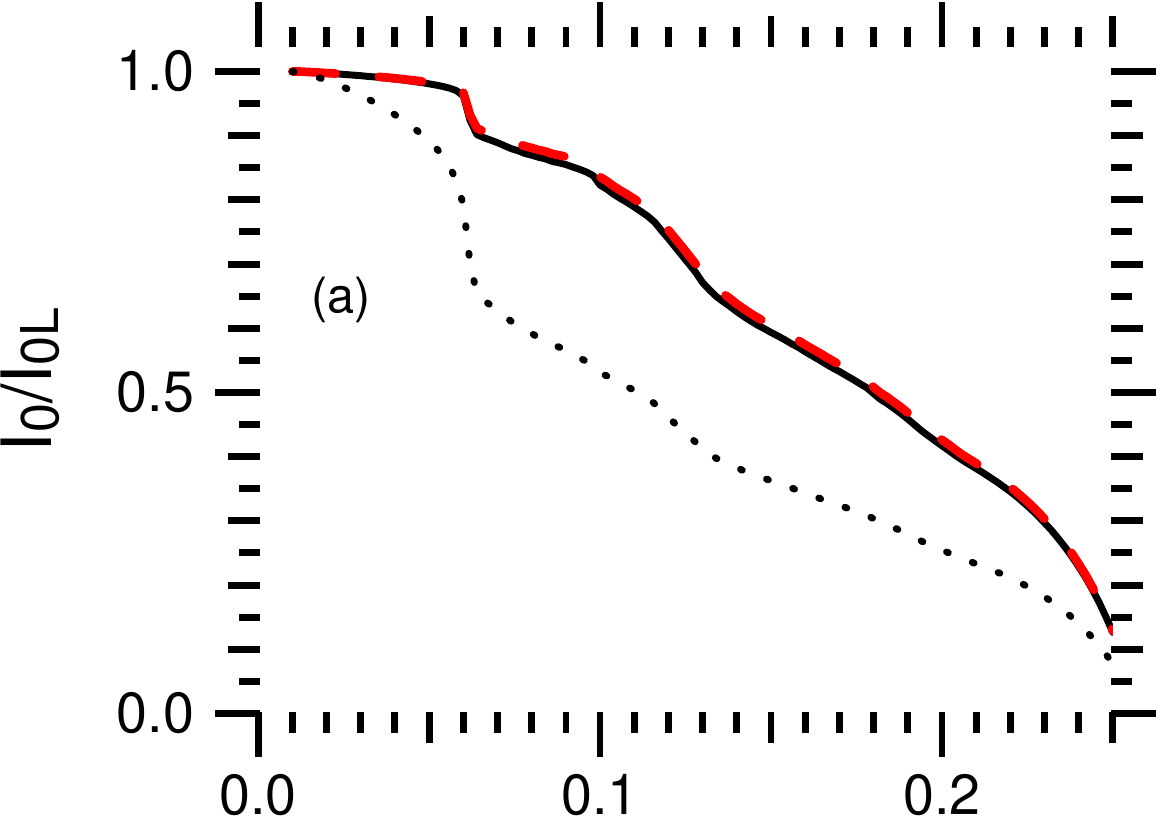} \\
  \includegraphics[width=2.4in]{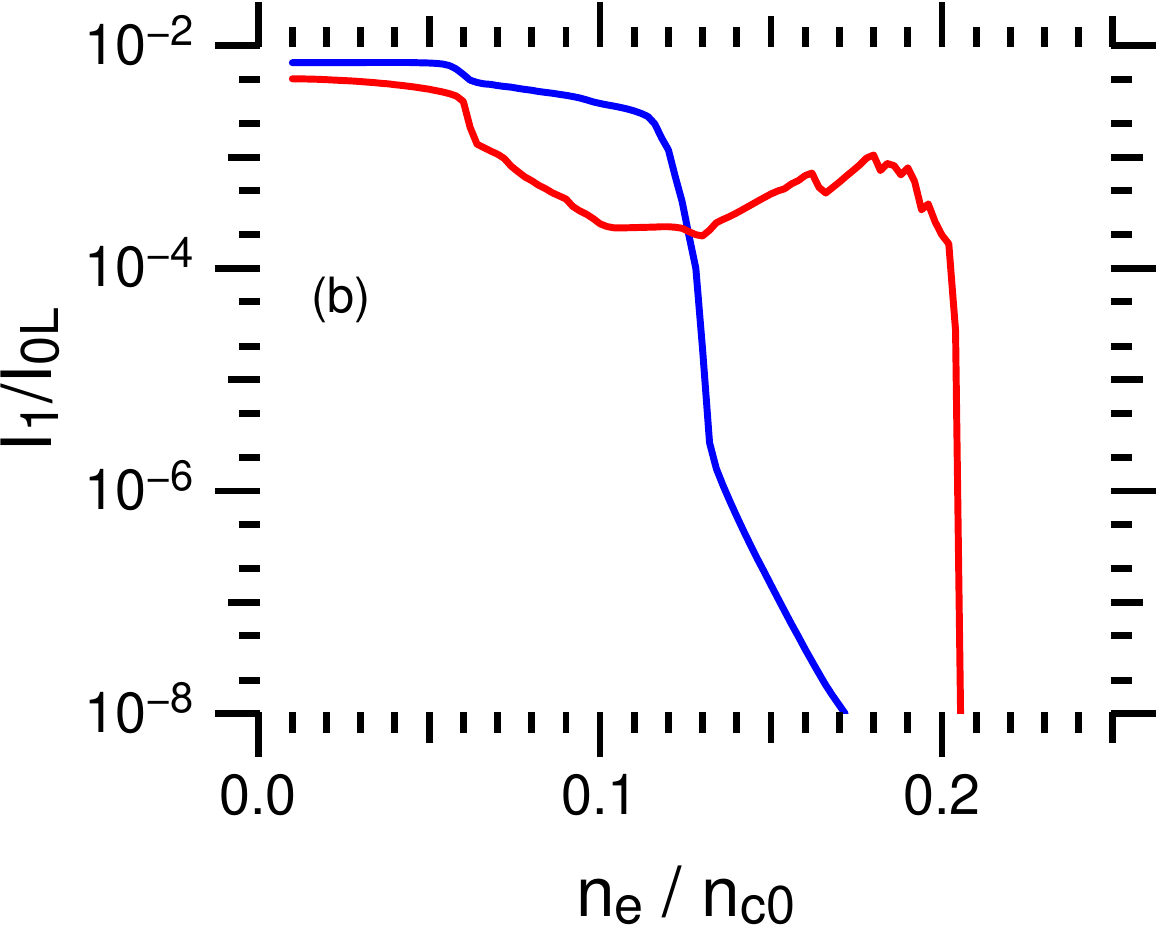}
  \caption{\colonl{} (a) Laser transmission for NIF example at 12.5 ns
    (peak power): black solid curve is the nominal \dep{} solution
    with pump depletion, red dashed curve is with just
    inverse-bremsstrahlung absorption, and black dotted curve is the
    \dep{} solution with $2\times\Ga_1$. (b) SBS (blue) and SRS (red)
    scattered intensities for the nominal \dep{} solution. Calculation
    of intensity at a given $n_e$ is described in text.}
  \label{f:niftrn}
\end{figure}

\begin{figure}
  \includegraphics*[width=2.75in]{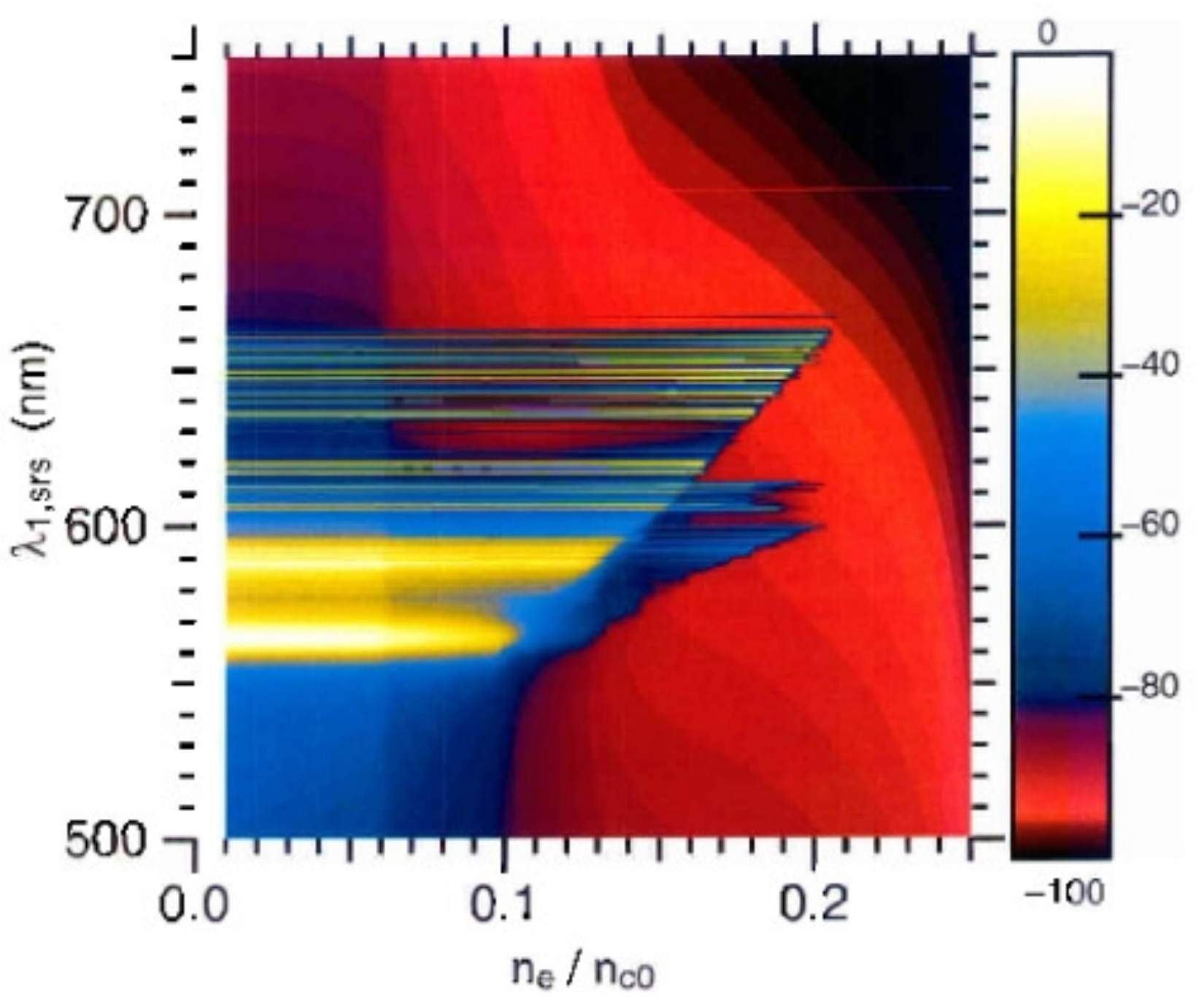}
  \caption{\colonl{} SRS spectral density $i_1$ vs.\ $n_e/n_{c0}$ and $\la_1$, in decibels, at 12.5 ns (peak power), for NIF example.}
  \label{f:nifiofsr}
\end{figure}

\begin{figure}
  \includegraphics*[width=2.75in]{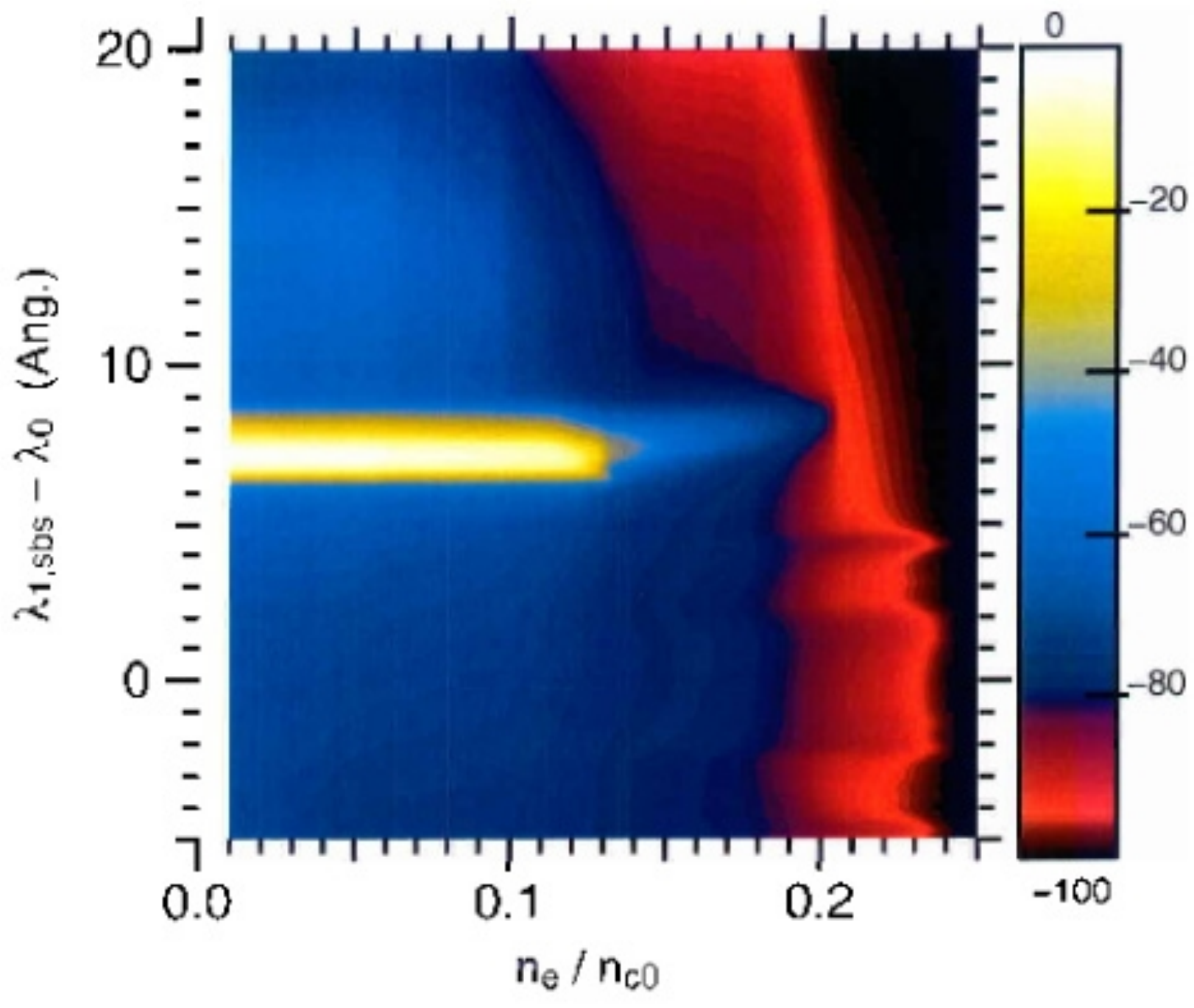}
  \caption{\colonl{} SBS spectral density $i_1$ vs.\ $n_e/n_{c0}$ and $\la_1-\la_0$, in decibels, at 12.5 ns (peak power), for NIF example.}
  \label{f:nifiofsb}
\end{figure}

Besides backscatter, \dep{} also provides the pump intensity $I_0(z)$
along each ray. This indicates how much laser energy is transmitted to
a given location, which is a crucial aspect of a whether LPI degrades
target performance. In cases where the backscattered light undergoes
significant absorption as it propagates out of the target (as happens
to SRS for the design analyzed here), the measured reflectivity can
understate the level of LPI. The laser transmission can reveal this
fact. Figure \ref{f:niftrn}(a) presents $I_0$, averaged over all the
rays, at a given $n_e$. This is a 1D presentation of how much energy
reaches a given density, although in the full 3D geometry
different rays reach the same $n_e$ at different locations. $I_0$ with
just pump absorption, as well as the \dep{} solutions with pump
depletion for the nominal case and $2\times\Ga_1$, are shown. Pump
depletion is barely discernible in the nominal case, but is
significant in the $2\times\Ga_1$ case. For instance, in the latter
case $I_0$ at $n_e/n_{c0}=0.2$ is only 60\% of its absorption-only
value. The wavelength-integrated SRS and SBS $I_1$ are shown in Fig.\
\ref{f:niftrn}(b), and the scattered spectra vs.\ $n_e$ are shown in Figs.\ \ref{f:nifiofsr}-\ref{f:nifiofsb}. SRS in particular develops at several different
densities, corresponding to different wavelengths, as can be seen in Figs.\ \ref{f:nifisrs} and \ref{f:nifiofsr}.

\section{Conclusions and future prospects} \label{s:conc}

We have derived a 1D, steady-state, kinetic model for Brillouin and Raman
backscatter, that includes pump depletion,
bremsstrahlung damping and fluctuations, and Thomson scattering. This
model is implemented by the code \dep{}, which we have presented as
well. This work extends linear gain
calculations, by including more physics while retaining its low
computational cost. In particular, \dep{}
provides the scattered-light spectrum and intensity developing from
physical noise, which can be compared against more sophisticated codes
and experiments. The transmitted pump laser along the profile is also
found, which is important for assessing an ICF target design,
especially when re-absorption of scattered light reduces the escaping
backscatter from its internal level.

We presented benchmarks of \dep{} on contrived, linear profiles, as
well as analysis of OMEGA experiments and a NIF ignition design. The
benchmarks reveal the deficiencies of linear gain, namely the neglect
of TS, pump depletion, and re-absorption. Comparisons with \ftd{}
provide a cross-validation of the two codes in a regime where they
should agree. The OMEGA SBS experimental data, as well as \ftd{}
simulations of these shots, show much more reflectivity than \dep{}
gives, for intensities where pump depletion is weak. This enhancement
is due to speckle effects. We showed an upper bound on this
enhancement is given by doubling the \dep{} coupling coefficient
$\Ga_1$, which comes from considering the phase-conjugated mode in an
RPP-smoothed beam. The ignition design analysis gives reasonably low
backscatter levels for the nominal laser intensity and including
re-absorption, with SRS dominating SBS. However, if re-absorption is
neglected, or especially if $\Ga_1$ is doubled, the backscatter
appears more worrisome. The laser transmission supports these
conclusions.

Ray-based gain calculations have been used for some time to model LPI
experiments, and \dep{} can provide more detailed comparisons.  An
early application of gain to hohlraum targets is Ref.\
\cite{glenzer-smoothing-pop-2001}, where hohlraums filled with CH gas
were driven by laser beams with and without PS and SSD.  Without SSD,
reasonable agreement was found between measurments and the time-dependent
SBS gain spectrum. However, there was a large difference in peak SRS
wavelength between measurements and the gain spectrum, which may be
due to laser filamentation changing the location of peak SRS growth.

Several future directions exist for \dep. One is to include an
``independent speckle'' model for gain enhancement, where one solves
the \dep{} equations over a speckle length for a distribution of pump
intensities and then re-distributes the power. This would not describe
correlations among axial ranks of speckles, caused e.g.\ by phase
conjugation. \dep{} also enables some new diagnostics and
applications. The pump and scattered intensities found by \dep{} can
be used to compute the local material heating rate due to
absorption. This could be incorporated into a hydrodynamic code,
thereby coupling LPI to target evolution in a self-consistent, if
simplified, way. In addition, the plasma-wave amplitudes found by
\dep{} can be compared against thresholds for various nonlinearities
to assess their relevance, and may allow estimation of hot electron
production by SRS.

Despite its promise, there are limits inherent to any 1D or ray-based approach,
stemming from 3D wave optics (e.g. diffraction, speckles,
filamentation, and beam bending). A 3D paraxial code called \slip{}
\cite{froula-lengthlim-prl-2008}, which like \dep{} operates in steady
state and uses kinetic coefficients, is being developed. This model is
in some sense intermediate between \dep{} and \ftd. 1D codes
like \dep{} still have a valuable role.  They can analyze hundreds of
rays, using hundreds of scattered wavelengths, in $\sim$ minutes, thus
allowing designs to be rapidly analyzed and compared. The resulting
time-dependent spectra allow for contact with experimental
diagnostics, and are frequently needed, for example, to choose the
carrier $k$ and $\om$ for \ftd.

Laser-plasma interactions have proven to be a very challenging area of
plasma physics, owing to the variety of relevant physics and extreme
range of scales involved. This has led to an equally extreme range of
modeling tools, from 1D gain estimates to 3D kinetic
simulations. By fully exploiting these tools, each with their uses and
limitations, a more complete picture is emerging.

\begin{acknowledgments}
  We gratefully recognize A.\ B.\ Langdon, R.\ L.\ Berger, C.\ H.\ Still, and L.\ Divol for
  helpful discussions and support. This work was supported by US
  Dept.\ of Energy Contract DE-AC52-07NA27344.
\end{acknowledgments}

\appendix
\section{\lip}
In this appendix, we document the laser interaction post-processor
\lip{}, of which \dep{} can be viewed as an extension. The equations
underlying \lip{} are
\begin{eqnarray}
  d_zI_0(z) &=& -\ka_0I_0,   \label{eq:lip1} \\
  \p_zi_1(z,\om_1) &=& -I_0\Ga_1i_1.  \label{eq:lip2}  
\end{eqnarray}
The first of these is Eq.\ (\ref{eq:I0gov}) with no pump depletion
($\tau_1=\Ga_1=0$), and the second is Eq.\ (\ref{eq:i1gov}) with no
bremsstrahlung or TS ($\ka_1=\Sigma_1=\tau_1=0$). That
is, only the absorption of the pump, and coherent coupling to
scattered light waves, are modeled. The boundary conditions are
$I_0(z=0)=I_{0L}$ (the known pump at the laser entrance), and
$i_1(z=L_z,\om_1)=1$. We thus solve for a unit scattered-wave boundary
seed, which is permissible for this linear system.

We readily solve Eqs.~(\ref{eq:lip1}-\ref{eq:lip2}) to find
\begin{eqnarray}
  \label{eq:1}
  I_0(z) &=& I_{0L}e^{-\int_0^z dz'\ \ka_0(z')}, \\
  i_1(z) &=& e^{G_l(z)}, \\
  G_l(z)   &\equiv& \int_z^{L_z} dz'\ \Ga_1(z')I_0(z').
\end{eqnarray}
$G_l(z)$ is the linear intensity gain exponent, and is \lip{}'s main
result. The total gain across the profile is $G_l(z=0)$.

The numerical computation of $G_l$ suffers from the problem of narrow
resonances, similar to \dep{}. The coupling coefficient $\Ga_1$ (see
Eq.~(\ref{eq:Gam1res})) is sharply peaked near the resonance point where
$\re\,\ep=0$. \lip{} addresses this challenge in a way analogous to
how \dep{} handles the coupling-Thomson step, as outlined in Appendix
B. In particular, the integration of Eq.~(\ref{eq:lip2}) from $z^n$
down to $z^{n-1}$ can be cast in the form
\begin{equation}
  \ln{i_1^{n-1} \over i_1^n}= \im\,S_0, \qquad S_0 \equiv \int_{z^n}^{z^{n-1}}dz\ {S \over \ep},
\end{equation}
with $S\equiv -I_0f\Ga_S\chi_e(1+\chi_I)$. Although $S(z)/\ep(z)$ is
sharply-peaked near resonance, $S(z)$ and $\ep(z)$ are themselves
slowly-varying with $z$. We approximate $S(z) \approx
S^{n-1/2}+(z-z^{n-1/2})\De S/\De z$ (and similarly for $\ep$), with
$X^{n-1/2} \equiv (X^n+X^{n-1})/2$ and $\De X \equiv (X^n-X^{n-1})$
for some quantity $X$ . With this representation, and $\hat X\equiv X^{n-1/2}/\De X$, we find
\begin{equation}
  S_0 = {\De z \De S \over \De \ep} \lb 1 + (\hat\ep-\hat S) \ln{\ep^{n-1}\over \ep^n} \rb.
\end{equation}
This formula is valid provided either $|\re\, \hat\ep| \geq 1/2$ or $\im\, \hat\ep\neq0$.
For accuracy, we also want $|\De\ep|$ to not be too small (which obtains, e.g., for a flat profile). We therefore require $|\hat\ep|$ to be less than some large number. If any of these conditions does not hold, we simply assume
$S=S^{n-1/2}$ and $\ep=\ep^{n-1/2}$ across the cell to find
\begin{equation}
  S_0 = \De z{S^{n-1/2} \over \ep^{n-1/2}}.  
\end{equation}

\section{Numerical solution of the coupling-Thomson step}
This appendix provides a derivation of Eq.\ (\ref{eq:i1CTsol}), the
solution for $i_1$ in the coupling-Thomson step. We must solve Eq.\
(\ref{eq:coup1}), from $z^n$ down to $z^{n-1}$, with $I_0=I_0^n$ and
all coefficients \textit{except} $|\ep|^2$ evaluated at
$z^{n-1/2}$. We write this equation as
\begin{eqnarray}
  \p_zi_1 &=& -  {K_\tau+K_\Ga i_1 \over |\ep|^2}, \label{eq:Ai1gov} \\
  K_\tau  &\equiv& I_0^n\tau_S^{n-1/2}g_\tau^{n-1/2}, \quad  
  K_\Ga   \equiv fI_0^n\Ga_S^{n-1/2}g_\Ga^{n-1/2}.
\end{eqnarray}
As mentioned above, the principal numerical difficulty is that
$|\ep|^{-2}$ is sharply peaked near resonance ($\re\, \ep=0$). Since
$\re\, \ep$ generally passes through zero slowly, we
Taylor expand $\ep$ within each zone and solve the resulting system
analytically.

Define the zonal average and difference $X^{n-1/2}\equiv
(1/2)(X^n+X^{n-1})$ and $\De X \equiv X^n-X^{n-1}$ for the quantity
$X$. We expand $\ep$ about the zone center $z^{n-1/2}$ to find
\begin{eqnarray}
  \ep &\approx& \ep^{n-1/2} + \hat z \De\ep, \label{eq:eptaylor} \\
  \hat z &\equiv& {z-z^{n-1/2}  \over \De z}.
\end{eqnarray}
We can then write
\begin{eqnarray}
  |\ep|^2 &=& \ep_1 + |\De\ep|^2(\hat z - \hat z_0)^2, \\
  \ep_1 &\equiv& |\De\ep|^{-2} \im\lb \ep^{n-1/2}\De\ep^*\rb^2 , \\
  \hat z_0 &\equiv& -|\De\ep|^{-2} \re\lb \ep^{n-1/2}\De\ep^*\rb .  
\end{eqnarray}
The linear change of variable
\begin{equation}
  s \equiv {|\De\ep| \over \sqrt{\ep_1}}(\hat z - \hat z_0)
\end{equation}
transforms Eq.~(\ref{eq:Ai1gov}) to
\begin{eqnarray}
  \p_si_1 &=& -{B_\tau+B_\Ga i_1 \over 1+s^2}, \\
  B_{\tau,\Ga} &\equiv& {K_{\tau,\Ga}\De z \over |\im\lb \ep^{n-1/2}\De\ep^*\rb|}.
\end{eqnarray}
A second change of variable to $w\equiv\atan\ s$ yields
\begin{equation}
  \p_wi_1=-(B_\tau+B_\Ga i_1).  
\end{equation}
This equation is solved to give the result used in Eq.~(\ref{eq:i1CTsol}):
\begin{eqnarray}
  i_1^{n-1} &=& (i_1^n+i_\tau)e^{B_\Ga\De w_n}-i_\tau, \\
  \De w_n &\equiv& \atan\ s^n - \atan\ s^{n-1}.  
\end{eqnarray}
$i_\tau=B_\tau/B_\Ga$ is also given by Eq.\ (\ref{eq:itau}).

If $\De\ep$ is sufficiently small (or zero, as for a flat profile), we
do not use Eq.\ (\ref{eq:eptaylor}), but instead
$\ep\approx\ep^{n-1/2}$. We can then immediately solve Eq.\ (\ref{eq:Ai1gov}) to find
\begin{equation}
  i_1^{n-1} = (i_1^n+i_\tau)\exp\lb {K_\Ga|\ep^{n-1/2}|^{-2} \De z} \rb -i_\tau.    
\end{equation}



\end{document}